\documentstyle[aaspp]{article}

\newcommand{\bq}{\begin{equation}}
\newcommand{\eq}{\end{equation}}

\begin{document}
\def\refitem{\par\parskip 0pt\noindent\hangindent 20pt}
 
\title{The Risetime of Nearby Type Ia Supernovae}

\author{Adam G. Riess$^1$, Alexei V. Filippenko$^1$, Weidong Li$^1$, Richard R. Treffers$^1$, Brian P. Schmidt$^2$,
Yulei Qiu$^3$, Jingyao Hu$^3$, Mark Armstrong$^4$, Chuck Faranda$^5$, Eric Thouvenot$^6$}

\affil{$^1$Department of Astronomy, University of California, Berkeley, CA 94720-3411,
$^2$Mount Stromlo and Siding Spring Observatories, Private Bag, Weston Creek, ACT 2611, Australia, $^3$Beijing Astronomical Observatory, Chinese Academy of Sciences,
Beijing 100080, China, $^4$UK Supernova Patrol, British Astronomical Association, Rolvenden, England, $^5$ 7860 NW 53 CT., Lauderhill, FL  33351, $^6$ Pic du Midi Observatory, France}

\begin{abstract}

    We present calibrated photometric measurements of the earliest detections
of nearby type Ia supernovae (SNe Ia).  The set of $\sim$30 new, unfiltered CCD observations delineate the early rise
behavior of SNe Ia $>$ 18 to 10 days before maximum.  Using simple empirical
models, we demonstrate the strong correlation between the risetime (i.e., the time between 
explosion and maximum), the post-rise light-curve shape, and the peak luminosity.  Using a
variety of light-curve shape methods, we find the risetime to $B$ maximum for a SN Ia
with $\Delta m_{15}(B)$ =1.1 mag and peak $M_V$=$-$19.45 mag to be 19.5$\pm0.2$ days.  We find that the peak brightness of SNe Ia is correlated with their risetime; SNe Ia which are 0.10 mag brighter at peak in the $B$-band require 0.80$\pm0.05$ days longer to reach maximum light.  We determine the effects of several possible sources of systematic errors, but none of these
significantly impacts the inferred risetime.    Constraints on SN Ia progenitor systems
 and explosion models are derived from a comparison between the observed and
 theoretical predictions
of the risetime.

\end{abstract}
subject headings:  supernovae: general$-$cosmology: observations

\vfill
\eject
 
\section{Introduction}

  A few weeks after explosion the visual
luminosity of a type Ia supernova (SN Ia) increases a
trillion-fold, by which time its peak output
 can rival the glow of its host galaxy.  Unfortunately,
 this dramatic rise to prominence is difficult to detect and therefore remains poorly documented
 (Vacca \&
Leibundgut 1996; Leibundgut et al. 1991a,b).   Enough of this rise occurs in the first few hours after explosion that it is possible
to detect low-redshift
 ($z < 0.02$) SNe Ia which are less than a day old.

  Discovering SNe Ia in their youth 
requires great persistence and good fortune; potential host galaxies
must be monitored frequently to increase the odds of an early detection.  Even when SNe Ia are detected
early in their development,  the observations are often recorded
 with a medium that does not easily lend itself to precise, quantitative analysis: naked-eye
 observations, photographic plate images, and unfiltered CCD images.
  To facilitate comparisons to subsequent observations, early SN Ia observations need to be reliably calibrated on standard passband systems using linear detectors.

  To date, the earliest
 reliable and precisely quantifiable detection (i.e., employing a CCD and a standard passband) of a nearby SN Ia 
is $\sim$ 13 days before
 $B$-band maximum, but still $\sim$ 1 week after explosion
 (SN 1994D: Richmond et al. 1995).   Detections of SNe Ia
 earlier than 10 days before $B$ maximum have been reliably measured
 and reported
 for only 4 nearby SNe Ia (SN 1994ae: Riess et al. 1999a; SN 1994D:
 Richmond et al. 1995; SN 1992bc: Hamuy et al. 1996; SN 1990N: Lira et
 al. 1998, Leibundgut et al. 1991a).  This is unfortunate, as the SN Ia rise to glory during 
the time interval
between explosion and maximum brightness (hereafter referred to as the
``risetime'') holds significant clues for understanding the
progenitors and physics of SNe Ia (Leibundgut \& Pinto 1992) and
 ultimately their utility for the determination
of cosmological parameters.  

\subsection{A Constraint on SN Ia Progenitors}

Precise knowledge of the SN Ia risetime could provide
valuable constraints on models of SN Ia progenitors.
  Although the ways the explosion characteristics influence the 
SN Ia rise behavior are complex, requiring detailed
simulations to understand, some general dependences can be understood.

  The rise behavior of a SN Ia is determined by the rate at
which energy in the interior is released and subsequently diffuses to the surface
of the supernova.  Although the energy deposition rate is always
decreasing, homologous expansion of 
material continually increases the rate at which energy
diffuses to the surface during the rising phase.  At this epoch, the
photosphere grows in radius while receding through the expanding ejecta. 

 The
available energy source is the radioactive $^{56}$ (followed by radioactive $^56$Co) which is synthesized in the explosive burning of carbon and
oxygen to nuclear statistical equilibrium (Fowler \& Hoyle 1960; Arnett 1969; Colgate \& McKee 1969; 
Arnett
1982; H\"{o}flich, Khokhlov, \& Muller 1992; 
Leibundgut \& Pinto 1992; Pinto \& Eastman 1999).  The initial progenitor mass plays an
 important role,
providing both the source of fuel and the obstruction to energy diffusion.
  The rate of diffusion to the
surface depends on the proximity of the radioactive
 material to the photosphere, the temperature, and the density of the homologously expanding envelope, factors which determine the opacity of the supernova.

  Detailed
modeling of the explosion yields predictions of the dependence of the
progenitor initial conditions and explosion mechanism on the SN Ia
risetime (H\"{o}flich \& Khokhlov 1996; H\"{o}flich, Wheeler, \& Thielemann
1998).   In general, single white dwarf progenitor systems seem to produce the 
shortest risetimes.  The risetime of double degenerate systems is extended by the additional barrier
to diffusion presented by the thick disk of the tidally disrupted companion.   Among single
degenerate systems with a mixture of deflagration and detonation
burning fronts, explosions with larger fractions of supersonic burning  produce
shorter risetimes (H\"{o}flich \& Khokhlov 1996).  

\subsection{A Theoretical Calibration of SN Ia Luminosity}

   Knowledge of the SN Ia risetime also provides the means to
calibrate the peak luminosity of SNe Ia
independent of the Cepheid distance scale.  One method is to use the
risetime to indicate the instantaneous rate of energy deposition from
the radioactive decay of $^{56}$Ni and $^{56}$Co.  This method still
relies on theoretical modeling to indicate the likely mass of
decaying nickel and the relation at maximum between the energy deposited
and radiated (once known as Arnett's Law: Arnett 1982; Arnett, Branch, \& Wheeler 1985).
  An alternate
route is to use the risetime to calibrate the luminosity of 
the expanding SN Ia photosphere at peak.  This method also has theoretical
input in the form of a spectral synthesis model.  A comparison of the
model and observed spectrum at peak luminosity yields the velocity of the
photosphere and its model temperature.   Together with the risetime, this information 
yields the size of the photosphere and hence the luminosity of the supernova at maximum.
  Although both methods are
only semi-empirical, each is essentially independent, relying on a
different aspect of SN Ia theoretical modeling.  To be consistent, the
two methods require a risetime to bolometric peak of
14 to 22 days (Nugent et al. 1995).

  Because measurements of the risetime provide a new window into the physics
  of SNe Ia, a comparison of the risetime for high-redshift and low-redshift
  SNe Ia could be a valuable test of evolution.  Such a comparison is
  performed by Riess et al. (1999b).  

Here we report calibrated measurements of the earliest 
detections of low-redshift SNe Ia and the risetime
inferred from them.  These observations, beginning fewer than
two days after the explosion, provide new constraints on SN Ia
evolution, progenitors, and luminosity. 
  As discussed in \S 2, we have collected the earliest CCD images of
SNe Ia.  These images are usually unfiltered and we have transformed
them to standard passbands.  We analyze
the SN Ia rise behavior in \S
3.  In \S 4 we explore and analyze possible systematic errors and biases in our measurements.
  We discuss the implications for SN Ia physics and progenitors in \S
5.

\section{Observations}

\subsection{Sample Selection}

   To measure the SN Ia rise behavior we have collected the earliest available CCD images of SNe Ia.   To
compile our sample of observations of young SNe Ia the following
selection criteria were used:

\noindent (1)  SNe Ia with well-sampled CCD light curves. \\
(2)  SNe Ia whose hosts were observed with CCDs between 10 and 25 days before $B$ maximum. \\
(3)  Spectroscipically normal SNe Ia (Branch, Fisher, \& Nugent 1993).\\
(4)  SNe Ia with redshift $z \leq 0.02$.  \\

  We limited our analysis to CCD observations because of their well known advantages: they are highly linear detectors over a large dynamic range and their digital output allows us to accurately subtract contamination from underlying host galaxy light.
Criterion (1) was necessary to be able to accurately determine the date
of $B$ maximum and therefore the age of all observations.
Criterion (2) targets the observations of SNe Ia which will yield
the most useful constraints on the rise behavior.  The period earlier
than 10 days before $B$ maximum was
chosen because this is the epoch where exceedingly few
SNe Ia have been observed and where new data are needed (see \S 1). 
Criterion (3) was chosen so that the inferred rise behavior will be
representative of prototypical SNe Ia.  Peculiar SNe Ia have been shown to
deviate substantially from linear relations derived from typical SNe Ia.  
This requirement can and should
be relaxed in the future when a sizeable sample of very early observations of peculiar SNe Ia
has been collected.  Criterion (4), that the SNe Ia be nearby, is important so that observations at very early times can yield significant
constraints.  For example, at $z=0.02$ a SN Ia at $\sim$ 15 days before $B$ maximum will have an apparent brightness of $\sim$18.5 mag in $B$.  This also results in a measured rise
behavior which is representitive of SNe Ia at low redshifts, a
desirable characteristic for the comparison to the rise behavior of
SNe Ia
 at high
redshift (Goldhaber 1998; Goldhaber et al. 1999; Nugent 1998).

Very early detections of supernovae are most common in supernova searches which monitor potential host galaxies with high frequency.  An additional requirement is that past observations are
catalogued and
 stored so
that images of host galaxies obtained before SN 
discoveries can be retrieved for later analysis.  In most cases, SN
light in very early images is not recognized until later when the SN brightens.   

Two prolific discoverers of
supernovae  
are the Beijing Astronomical Observatory Supernova Search employing a 0.6 m telescope (BAO; Li et al. 1996) and 
the Lick Observatory Supernova Search (LOSS) with the 0.75 m Katzman Automatic Imaging Telescope (KAIT; Treffers et al. 1997; see also Richmond, Treffers, \& Filippenko 1993).  
The BAO search has been operating since April 1996 
 and to date has discovered 13 SNe Ia.  The LOSS has been
underway since the end of 1997 and to date has
discovered 14 SNe Ia.  

For all SNe Ia known to pass criteria (1), (3), and (4), we queried
the BAO and LOSS databases and obtained all host galaxy images
expected to be between 10 and 25 days
 before the observed $B$
maxima.  These observations are listed in Table 1 and include SNe
1996bo, 1996bv, 1996by, 1997bq, 1998dh, 1998ef, and SN 1998bu.

\begin{table}[t]
\begin{small}
\begin{center}
\vspace{0.4cm}
\begin{tabular}{lllllll}
\multicolumn{7}{c}{Table 1: SN Ia Early Detections} \\
\hline
\hline
SN Ia & $z$ & discovery$^a$ & date $B_{max}^a$ & date observed$^a$ & filter & source \\
\hline
1990N$^b$  & 0.003 & 8065 & 8082 & 8065-8072 & unfiltered &
E. Thouvenot \\
1994D & 0.001 & 9419 & 9432 & 9420, 9421 & $B, V$ & Richmond et al. (1995) \\
1996bo & 0.017 & 10379$^c$ & 10386 & 10375 & $V$ & BAO \\
1996bv & 0.017 & 10400$^d$ & 10404 & 10390 & unfiltered & BAO \\
1996by & 0.014 & 10432 & 10441 & 10432, 10429 & unfiltered & BAO \\
1997bq & 0.009 & 10546 & 10556 & 10541 & unfiltered & BAO \\
1998aq & 0.004 & 10916 & 10931 & 10910, 10916 & unfiltered & M. Armstrong \\
1998bu & 0.003 & 10943 & 10952 & 10932, 10936 & unfiltered & LOSS, C. Faranda \\
1998dh & 0.009 & 11018 & 11028 & 11010, 11014, 11018 & unfiltered & LOSS \\
1998ef & 0.018 & 11104 & 11114 & 11096, 11100, 11104 & unfiltered & LOSS \\
\hline
\hline
\multicolumn{7}{l}{$^a$ Julian Date$-$2,440,000.} \\
\multicolumn{7}{l}{$^b$ Also Lira et al. (1998) $B$,$V$ on
JD=8071, 8072.} \\
\multicolumn{7}{l}{$^c$ Date of independent discovery from
Armstrong et al. (1998).} \\
\multicolumn{7}{l}{$^d$ Date of discovery (Li 1998).} \\ 
\end{tabular}
\end{center}
\end{small}
\end{table}

  We also combed the IAU Circulars and an amateur supernova finders'
 network, VSNET (http://www.kusastro.kyoto-u.ac.jp/vsnet/SNe/SNe.html), for
 reports of very early observations of SNe Ia which also passed
 criteria (1), (3), and (4).  This search yielded
 two valuable additions: SN 1998aq and SN 1990N.  In addition, a pre-discovery detection of SN 1998bu in 
NGC 3368 (M96) was reported on VSNET by amateur astronomer C. Faranda, using an unfiltered CCD
 and 0.25 m telescope from his backyard in Lauderhill, Florida.  This
 nearby SN Ia was discovered approximately nine days before $B$
 maximum, but the image by Faranda was obtained 16 to 17 days before
 $B$ maximum (Meikle et al. 1998).  NGC 3368 was also observed by
 LOSS 4 days earlier (i.e., 20 to 21 days before $B$ maximum), though SN 1998bu was not detected.  
SN 1998aq was discovered by
 M. Armstrong (Armstrong et al. 1998) with a 0.26 m reflector and an unfiltered CCD 
in the course of the UK
 Supernova Patrol approximately 15 days before $B$ maximum, and the host
 galaxy was also imaged $\sim$ 21 days before $B$ maximum.  SN 1990N was discovered by
 E. Thouvenot (Maury et al. 1990) using a 0.6 m telescope and an unfiltered CCD 
about 17 days before $B$ maximum
 and was subsequently observed with the same equipment for the next seven
 days.  In all cases, the observers furnished us with
 their unfiltered CCD images (see Table 1). 

 We have included SN 1994D using the observations by Richmond et
 al. (1995) in $B$ and $V$ commencing $\sim$13 days before $B$
 maximum.  Although SN 1994ae was observed in $R$ about 13 days
 before $B$ maximum by the Leuschner Observatory Supernova Search
 (the predecessor to the Lick search; Richmond et al. 1994), Ho et al. (1999) have found that the Leuschner
 $R$ passband is a poor match to Johnson/Cousins $R$ and its
 transmission function has not been adequately quantified.  Hence
 we have not included SN 1994ae in the present analysis.  In \S 4 we address
 possible biases on the risetime measurements attributable to our
 sample selection.  

   A remarkably early detection of SN 1989B was reported by Marvin
 \& Perlmutter (1989) on behalf of the Berkeley Automated Supernova
 Search 17 to 18 days before $B$ maximum at approximately mag
 17.  Unfortunately,
 the high surface brightness of the knotty spiral arm ($R$=15.6 in a 7.4 arcsecond aperture at the position of the SN; Wells et al. 1994) makes it
 extremely difficult to evaluate the brightness of a $\sim$ 17 mag
 stellar object without the aid of sophisticated galaxy subtraction
 procedures.  Our inability to obtain this CCD image forces us to
 regard this early detection as merely anecdotal. 

 We note that several very early photographic SN Ia detections have also been
 reported.  One detection in this category was presented by
 Barbon et al. (1982) for SN 1979B in NGC 3913.  A pre-discovery
 photographic plate revealed the supernova at 4.5 to 5.5 mag
 below peak about 14 to 18 days before maximum.  
 The precision of this measurement is limited by the poor sampling of
 the light curve and 
 the difficulty in evaluating the accuracy of the photographic
 magnitudes.  In the course of monitoring SN 1980N in NGC 1316, Hamuy
 et al. (1991) obtained pre-discovery photographic images of SN 1981D
 at 15 to 16 days before maximum.  Although they report the SN to
 have been 5 to 6 mag below peak at this epoch, their inability to
 reliably subtract the galaxy background ($\sim$ 3 mag brighter
 than the SN) makes this measurement difficult to interpret.
 These data (and others like it) fail our
 selection criteria and have not been included in our analysis.

\subsection{Photometric Calibrations}

Several amateur and professional supernova searches
 use unfiltered CCDs to increase
 their efficiency: an unfiltered CCD image can reach fainter
 magnitudes in less time than a filtered CCD. 

 Of the $\sim$30 observations made approximately 10 or more days before
 $B$ maximum,  over 80\% of the unique temporal
 samplings of the early rise are recorded with unfiltered CCDs.  The images
 which employed standard filters were obtained at the oldest ages of
 our sample (10 to 13 days before $B$ maximum) and yield the weakest
 constraints on the rise behavior.  

   If we wish to quantify the behavior of SNe Ia at the
 earliest observed times we need to make use of the available unfiltered CCD
 observations of young SNe Ia.   
 By either empirically characterizing a CCD's sensitivity
 or with explicit knowledge of its response function,
 it is possible to accurately transform an unfiltered SN Ia magnitude to a
 standard passband for comparison to subsequent filtered observations. 
The same method has been used by Schmidt et al. (1998) to transform supernova magnitudes observed
 with non-standard filters to a standard system.  
Here we will review the steps in the transformation.
 In what follows we will consider the standard passband to be
 Johnson $V$, but this
 transformation process is general and valid for other passbands.

  We first measured the differences in the unfiltered magnitudes between
  the SNe and individual local standard (LS) stars in the SN fields.  
In cases with significant host galaxy light at the position of the SN,
we initially subtracted template images of the host galaxy 
  and then fit point-spread functions to the SN and LS stars
 (Schmidt et al. 1998).  

  Referring to an individual, unfiltered CCD system as a non-standard
  passband system called ``$W$'' (for white light), the result is the measured quantity 
  \bq \Delta W=W_{SN}-W_{LS}, \eq
  or the magnitude difference between the SN and any LS star
  in the $W$ passband.

  On photometric nights at each of the telescopes, Landolt (1992) standard fields were observed
 to derive transformations between instrumental and expected magnitudes.  These transformations were used to
 calibrate Johnson $B$ and $V$ passband magnitudes and
 colors of the
 LS stars. 

  Now we seek a transformation of the LS stars from the standard
  passband, in this case $V$, to the non-standard, $W$.  This transformation is
  simply the pseudo-color, $V-W$, for
  any LS star.  If we know the spectral energy distribution (SED) of a LS star,
  this pseudo-color is

  \bq V-W=-2.5 \log \left({\int S_V(\lambda) F_\lambda(\lambda)d\lambda}{\int S_W(\lambda) d\lambda} \over
{\int S_W(\lambda) F_\lambda(\lambda)d\lambda}{\int S_V(\lambda)
d\lambda} \right)  + ZP_V-ZP_W, \eq where $F_\lambda(\lambda)$ is the spectrum of a LS star
 in units of power per unit area per unit wavelength, $ZP_V$ and $ZP_W$ are
 the zeropoints of the $V$ and $W$ filter systems\footnote[1]{Zeropoints are commonly defined to yield
 $V$(Vega)=$W$(Vega)=0.03, though this criterion is arbitrary here
since, as will be seen, the transformations are independent of
the choice of passband zeropoints.}, and $S_V(\lambda)$ and $S_W(\lambda)$ are the
 dimensionless sensitivity functions of the $V$ and $W$ systems.   

We refer to $V-W$ as defined by equation (2) as a
  ``pseudo-color'' because the sensitivity functions of the $V$ and
  $W$ bands generally have much more
overlap than do the passbands in most standard systems.

   For each observatory's $S_W(\lambda)$  we used the
 manufacturer's specified CCD response function (or equivalently the
 quantum efficiency divided by the wavelength) and multiplied by the atmospheric
 transmission (Stone \& Baldwin 1983).  In addition, we have taken care to
 include the effects of materials in the light paths of the telescopes.
 The most
 important changes in the true response function are in the near
 ultraviolet (i.e., 3000 \AA\ to 4000 \AA);  at these wavelengths, layers of
 glass in the camera lens or covering the CCD window can 
 affect the throughput of the telescope.   In addition, aluminum and
 other mirror coatings can affect the ultimate transmission.  Among
 the observatories in Table 1,  these effects are only
 significant for KAIT (LOSS) which boasts a strong UV CCD
 response.  Figure 1 displays the effective CCD response functions for
 the observatories listed in Table 1.  In Table 2 response
characteristics for these CCDs are listed.    Column 3 contains
 the wavelengths of peak efficiency ($\lambda_0$),  column 4 gives the wavelength range 
within which the CCDs efficiency is more than 10\% of the peak
 ($\lambda_{10\%}$), and column  5 lists the full width at half maximum sensitivity ($\Delta\lambda$). 

    For each unfiltered system in Table 2 we have numerically integrated equation
(2) using a set of 175 spectrophotometric standard stars from Gunn \&
Stryker (1983).  In Figure 2 we show the synthetic values of $V-W$ for these stars
versus the standard colors, $B-V$,  where the $W$ system
illustrated is from BAO.  The stellar SEDs result in a simple linear
relation between the standard color, $B-V$, and the pseudo-colors.   These theoretically derived
 relations demonstrate that such transformations are highly linear and have a
dispersion of less than 0.05 mag for stars with a color of $-$0.2 $ < B-V < $ 1.5.  By measuring a star's $B-V$ color we can use these relations to determine its
pseudo-color and equivalently its transformation to the $W$ system.

   We can now approximate equation (2) with a simpler
expression valid for stars:
\bq V-W=C_{VW}(B-V), \eq
where $C_{VW}$ is the slope of the linear relation between the
standard color and the pseudo-color and we impose the condition that
$V-W=0$ when $B-V=0$. 
While in principle we can derive the values of $C_{VW}$ from the
theoretical sensitivity functions of the $W$ passbands and spectrophotometry of stars,
 the most reliable way
to solve for these values is to empirically measure them from the
observed magnitudes of LS stars.
  This method has the advantage that it reduces our
reliance on accurately assessing the sensitivity functions of the $W$
systems.  

For all the LS stars in the fields of the SNe, we empirically fit the
values of $C_{VW}$ with equation (3) using the instrumental $W$ system
magnitudes of the LS stars.  In some cases we were able to obtain the $W$ system mags of
Landolt (1982) fields.
These $C_{VW}$ are given in Table 2 (column 6), as
are transformation coefficients to $B$ (column 7), $C_{BW}$.  
Some empirical insight into the response functions of the unfiltered
 CCD systems can be
gained by comparing the values of the coefficients, $C_{VW}$, to the 
correlation coefficients between two standard colors.  The relations
between the standard color,
$B-V$,
and other standard colors calculated from the Gunn \& Stryker (1983) spectrophotometric
standards are
\bq  V-I = 0.94(B-V),  \eq
\bq V-R = 0.51(B-V),   \eq
\bq V-V \equiv 0.0(B-V),  \eq
and \bq V-B \equiv -1.0(B-V), \eq
where the first two relations hold for stars with $B-V < 1.3$ mag.

As seen in Table 2, the values of $C_{VW}$ range from 0.3 to 0.5.
Comparing equation (3) to equations (4) through (7), we note that the
unfiltered CCD response functions have peak efficiency between $V$ and $R$, 
although the wavelength regions between $B$
and $I$ are also sampled as seen in Figure 1 and Table 2.
\begin{table}[t]
\begin{small}
\begin{center}
\vspace{0.4cm}
\begin{tabular}{lllllll}
\multicolumn{7}{c}{Table 2: Stellar Unfiltered Transformations} \\
\hline
\hline
Observer & CCD & $\lambda_0$ (\AA) & $\lambda_{10\%}$ (\AA) &
$\Delta\lambda$ (\AA) & $C_{VW}$($\sigma$) &  $C_{BW}$($\sigma$) \\
\hline
BAO & TI TC211 & 6800 & 3500-10000 & 4000 & 0.40(0.07) & 1.40(0.07) \\
KAIT (LOSS) & SITe UV2AR & 5000 & 3000-10000 & 5500 & 0.33(0.10) & 1.33(0.10) \\
E. Thouvenot & Thomson TH7863 & 7500 & 4000-10000 & 3500 & 0.40(0.11) & 1.40(0.11) \\
M. Armstrong & Sony ICX027BL & 5500 & 3700-9000 & 3500 & 0.31(0.07) & 1.31(0.07) \\
C. Faranda & SBIG ST6 & 5500 & 4000-10000 & 3700 & 0.51(0.09) &
1.51(0.09) \\
\hline
\hline
\end{tabular}
\end{center}
\end{small}
\end{table}
As shown in Figure 2, the theoretical and
empirical determinations of $C_{VW}$ are in good agreement.  This was
true for all $W$ systems in Table 2.  

  By use of equation (3) (the empirical version of
  equation (2)), we have transformed the magnitudes of the LS stars in
  $V$ (i.e., the standard system) to 
  the $W$ system.   Adding these transformed magnitudes
  to expression (1) yields the
  magnitude of a SN on the $W$ system.  

  Unfortunately, early SN magnitudes in 
these non-standard systems cannot readily be compared to subsequent observations through
standard passbands.  Therefore, we seek a transformation of the SN
 magnitude from the $W$ system to a standard passband system.
(An alternative is to transform subsequent SN magnitudes calibrated 
on standard systems to the $W$ system,
but the results would be difficult to evaluate.)

 The transformation we need is the inverse (i.e., the negative) of equation (2), 
except $F_\lambda(\lambda)$ is now the SED 
 of an infant SN Ia.  This pseudo-color is equivalent to a cross-filter
 $K$-correction between the $W$ and $V$ systems as expressed by
Kim, Goobar, \& Perlmutter (1996), except no redshifting
 of the spectrum is involved (i.e., $z=0$).  The appropriate SN Ia spectrum
 for this calculation is one at a similar age and color as the SN Ia
 to be transformed.  We employed the earliest spectrum of SN 1990N
 shown in Figure 1 (Leibundgut et al. 1991a) and
 the earliest spectrum of SN 1994D (Filippenko 1997), both
 obtained $\sim$ 14 days before $B$ maximum.  
These spectra should provide an accurate model of the SED because their age coincides with
the median age of our sample of SNe when they were observed.
 Because a few of the unfiltered CCD
 response functions have some minor sensitivity redward of 8200 \AA\
and blueward of 3600 \AA, spectra of SN 1994U (Riess et al. 1998a) and SN
 1990N (Leibundgut et al. 1991a) at 8 days before maximum were used to augment our description
 of the early SN Ia SED.  This augmentation was included for
 completeness but has very little effect on the evaluation of equation
 (2). 
 
  Nugent et al. (1999) have found that, to within 0.01 mag, the effects of both extinction and intrinsic color variation on the SN Ia SED can be reproduced by
application of a Galactic reddening law (Cardelli, Clayton, \& Mathis 1989).
  The augmented spectra of SN 1990N and SN
 1994D were reddened and dereddened using a Galactic reddening law 
to match the earliest
 $B-V$ color measurement of each SN Ia listed in Table 3 (column 2).  In all cases, the earliest
 color measurement occured 8 to 12 days before maximum light.

 Analysis of the early colors of SN 1990N (Leibundgut et al. 1991a),
 SN 1994ae (Riess et al. 1999a), SN 1994D (Richmond et al. 1995), SN
 1992bc (Hamuy et al. 1996a), SN 1997br (Li et al. 1999), SN 1995bd
 (Riess et al. 1999a), and
 SN 1998aq (Riess et al. 2000) demonstrates that SNe Ia evolve, on
 average, by less than 0.05 mag in $B-V$ between 10 to 14 days before $B$
 maximum.  We conservatively adopt a statistical uncertainty of 1$\sigma$=0.10
 mag for the individual $B-V$ color of a young SN Ia. If the SED of a SN Ia undergoes a dramatic (and
 unexpected) change between 16-18 days and 14 days before $B$
 maximum, a larger error could occur.   However, in \S 4.2 we
 demonstrate that this potential systematic error does not affect our analysis.
 The uncertainties resulting from intrinsic variation in the early SN Ia
 SEDs are calculated from the
 dispersion of a set of 12 SN Ia spectra in the range of 8 to 14 days before
 $B$ maximum.  

  Due to the breadth of the unfiltered response functions, we can reasonably
 transforming the data to either the $B$, $V$, or $R$
  passband systems.  The better the match between the
  filtered and unfiltered response functions, the less we must rely on
  the accuracy of the transformation.  If there
  is little overlap between the filtered and unfiltered response
  functions, the pseudo-color becomes similar to a ``conventional''
  color.  As seen in Figure 1,  the $V$ and $R$ passbands are the
  closest match to most of the unfiltered response functions.
  Yet there is a strong historical precedent for measuring
  risetimes relative to $B$ maximum.  In addition, for comparison to
  the risetime measured at high redshift, transforming to the $B$ passband
  is desirable (Goldhaber et al. 1999; Goldhaber 1998; Nugent 1998; Riess et al. 1999a). 
 For these reasons we have chosen to transform the data to
  both $B$ and $V$.  The greater uncertainty in the $B$ transformation
  is included in the individual photometric uncertainties.
 The values and uncertainties in the theoretical pseudo-colors,
  $W-V$ and $W-B$, are
 listed in Table 3 (columns 3 and 4, respectively).  Combining these with equation (1) yields
the SNe mags on the standard system.  We
  note that the $B$ and $V$ risetime measurements will not be
  independent.

  An alternate method to transform the unfiltered SN Ia magnitudes to
  a standard system is to employ the same empirical transformation previously used for
  stellar SEDs, equation (3), which only depends on the $B-V$ color of
  the SED.  It is well known that the SED of a SN Ia is extremely
 non-stellar over the vast majority of its observed lifetime (see
 Filippenko 1997 for a review).  However, at very early times the SED
  of a SN Ia can be photometrically approximated by a thermal continuum
 (see Figure 1), making the use of a
  transformation derived for stars plausible.  (It is likely that this thermal continuum
is even more dominant when the SNe Ia are younger than the earliest spectral observations.)
We used equation
  (3) and the values of $C_{VW}$ in Table 2 to derive SN Ia pseudo-colors,
  \bq W-V=-C_{VW}(B-V),\eq using the earliest measurements
  of the SN $B-V$ colors in Table 2.  These empirical SN pseudo-colors are listed
  in Table 3 (columns 5 and 6).  The empirical and theoretical pseudocolors
  are in good agreement with a relative dispersion of 0.08 and 0.11
  mag in $V$ and $B$, respectively.

  The advantage of using these completely empirical transformations
 is that they do not require explicit knowledge of
  the unfiltered response functions.  The advantage of using the theoretical SN pseudo-colors is that we avoid assuming that at very early times, SN Ia SEDs are
 thermal.  For the following analysis we will consider
  SN transformations derived from both empirical and theoretical
 pseudo-colors.  As 
 seen in Table 3 and in \S 3 and \S 4, the excellent agreement between the
 results derived from the two methods suggests that both methods
 are valid to better than $\sim0.1$ mag.

   For the unfiltered observations with higher signal-to-noise ratios ($S/N \geq 20$), the
   dominant source of uncertainty in the transformed standard
   magnitudes is the uncertainty in the determination of the SN
   pseudo-color listed in Table 3.  For the
   observations with lower signal-to-noise ratios ($S/N \leq 10$), the uncertainty is dominated by
   the statistics of the measured flux of the SNe.  

  We have performed a powerful 
cross-check of the transformation of 
unfiltered SN Ia magnitudes to a standard passband system by comparing the
results with independent, coeval magnitude measurements in
standard passbands.  We compared $B$ and $V$ magnitudes for SN 1990N
 at 11 and 12 days before $B$ maximum measured by Lira et al. (1998)
with those transformed to $B$ and $V$ 
using concurrent unfiltered observations of E. Thouvenot.
The mean difference was 0.03 and 0.01 mag for $V$ and $B$,
respectively (see Figure 3).  

  An even more
powerful test was performed by comparing complete $B$ and $V$ passband light
curves of SN 1997bq with $B$ and $V$ light curves transformed from
unfiltered measurements obtained by BAO.  In this case it was necessary
to use the theoretical SN Ia pseudo-colors derived from SNe
spectra at appropriate ages, since the
assumption that the SN Ia SED matches a thermal spectrum will obviously fail at
later times. The uncertainties in transformations at these later ages
are smaller due to the increased availability of spectra to model the SED.
As seen in Figure 3, the
agreement with the measurements employing passbands is excellent, with a dispersion of 0.03 and 0.04 mag in $V$
and $B$ (respectively) for 14 points.  

  We conclude that we can reliably transform unfiltered CCD SN Ia observations
  to a standard passband system at a valuable precisionof $\sim$5\%, and we expect that the
  resulting early measurements of SN Ia magnitudes can yield important
  constraints on the SN Ia risetime behavior.

\begin{table}[t]
\begin{small}
\begin{center}
\vspace{0.4cm}
\begin{tabular}{llllll}
\multicolumn{6}{c}{Table 3: SN Ia Transformations} \\
\hline
\hline
\multicolumn{2}{c}{} & \multicolumn{2}{c}{theoretical} & 
\multicolumn{2}{c}{empirical} \\
SN Ia & early $B-V$($\sigma$) & $W-V$($\sigma$) & $W-B$($\sigma$) & $W-V$($\sigma$) & $W-B$($\sigma$)  \\
\hline
SN 1996by & 0.46(0.03) & $-$0.13(0.11) & $-$0.58(0.11) & $-$0.18(0.04) &
$-$0.64(0.11) \\
SN 1996bv & 0.20(0.03) & $-$0.02(0.09) &  $-$0.21(0.10) & $-$0.08(0.04) &
$-$0.28(0.10) \\
SN 1997bq & 0.04(0.03) & 0.00(0.08) & 0.04(0.11) & $-$0.02(0.02) & $-$0.06(0.10) \\
SN 1998dh & 0.18(0.03) & 0.07(0.07) & $-$0.11(0.10) & $-$0.06(0.04) &
$-$0.25(0.11) \\
SN 1998ef & 0.10(0.03) & 0.06(0.07) & $-$0.04(0.12) & $-$0.03(0.03) &
$-$0.13(0.11) \\
SN 1998bu & 0.28(0.03) & $-$0.02(0.09) & $-$0.21(0.11) & $-$0.14(0.04) & $-$0.42(0.11)  \\
SN 1998aq & $-$0.18(0.03) & 0.02(0.05) & 0.20(0.11) & 0.06(0.03) & 0.24(0.11) \\
SN 1990N  & 0.05(0.03) & 0.05(0.10) & 0.00(0.11) & $-$0.03(0.04) &$-$0.08(0.10) \\
\hline
\hline
\end{tabular}
\end{center}
\end{small}
\end{table}

\section{Rise Curve}

   We first examined the SN Ia rise behavior by subtracting the fitted
   date and magnitude of maximum from each light curve.  The
   maxima were derived from second-order polynomial fits to data 
within five days of the peak.  (This is an iterative procedure.)  The
   individual data points minus the fitted peak magnitudes and dates are listed in Table 4.

\begin{table}[h]
\begin{small}
\begin{center}
\vspace{0.4cm}
\begin{tabular}{lllllll}
\multicolumn{7}{c}{Table 4: SN Ia Data} \\
\hline
\hline
SN & age$^a$ & $B$ mag$^b$ & $\sigma$(mag) & age$^c$ & $V$ mag$^d$ & $\sigma$(mag) \\
\multicolumn{1}{c}{} & \multicolumn{3}{c}{$B$} & \multicolumn{3}{c}{$V$} \\
\hline
SN 1996by&   $-$13.85 &       3.01 &      0.23 & $-$15.01 &       2.97 &      0.21\\
    &  $-$10.82 &       1.42 &      0.16 & $-$11.98 &       1.38 &
0.14\\
 &--- &--- &--- & $-$9.98 &      0.85 &     0.03\\
SN 1994D &      $-$11.71 &       2.35 &     0.06 & $-$12.67 &       2.12 &     0.05\\
     & $-$10.86 &       1.81 &     0.05 & $-$11.84 &       1.75 &     0.04\\
     & $-$9.90 &       1.39 &     0.05 & $-$10.86 &       1.36 &     0.04\\
SN 1997bq &  $-$15.67 &       4.01 &      0.18 & $-$18.63 &       4.26 &      0.16\\
SN 1998dh  &  $-$18.70 & $>$ 4.80 & 0.30 & -20.38 & $>$ 4.86 & 0.30 \\
& $-$14.70 &       3.01 &      0.14 & $-$16.38 &       3.04 &      0.12\\
    & $-$10.70 &       1.24 &      0.14 & $-$12.38 &       1.27 &      0.12\\
    & $-$9.70 &       1.09 &     0.03 & $-$11.38 &       1.12 &     0.03\\
&--- &--- &--- & $-$10.38 &      0.85 &     0.03\\
SN 1990N &     $-$17.92 &       3.32 &      0.15 & $-$19.98 &       3.33 &      0.13\\
  &   $-$17.02 &       2.74 &      0.15 & $-$19.08 &       2.75 &      0.13\\
  &   $-$16.02 &       2.24 &      0.15 & $-$18.08 &       2.25 &      0.13\\
  &   $-$15.02 &       1.97 &      0.15 & $-$17.08 &       1.98 &      0.13\\
  &   $-$14.02 &       1.56 &      0.15 & $-$16.08 &       1.57 &      0.13\\
  &   $-$13.02 &       1.34 &      0.15 & $-$15.08 &       1.35 &      0.13\\
  &   $-$12.02 &       1.12 &      0.15 & $-$14.08 &       1.13 &      0.13\\
  &   $-$10.92 &      0.87 &      0.15 & $-$12.98 &      0.87 &      0.13\\
  &   $-$11.92 &       1.11 &     0.03 & $-$13.98 &       1.12 &     0.03\\
  &   $-$10.92 &      0.86 &     0.03 & $-$12.98 &      0.91 &     0.03\\
 &--- &--- &--- & $-$9.98 &      0.42 &     0.03\\
 SN 1998bu & -20.74 & $>$ 6.30  & 0.30 & -22.04 & $>$ 6.50 & 0.30 \\ 
&   $-$16.74 &       4.65 &      0.18 & $-$18.04 &       4.75 &      0.15\\
 SN 1998aq & -20.82 & $>$ 5.13 & 0.30 & -22.08 & $>$ 5.25 &  0.30 \\
 &   $-$14.82 &       2.79 &      0.16 & $-$16.08 &       2.88 &      0.12\\
   &  $-$10.82 &       1.19 &     0.05 & $-$12.08  &    1.28  &  0.05\\
   &  $-$9.82 &      0.89 &     0.04 & $-$11.08  &   0.98  &  0.04\\
 &--- &--- &--- & $-$10.08  &   0.77 &   0.03\\
 SN 1998ef & $-$17.02 & $>$ 4.00 & 0.30 & $-$18.48 & $>$ 4.02 & 0.30 \\ 
 &    $-$13.02 &       2.32 &      0.14 & $-$14.48  &    2.27
& 0.12\\
 &--- &--- &--- & $-$10.48  &   0.80 &  0.12\\
 SN 1996bv &    $-$12.12 &       1.20 &      0.22 & $-$14.48  &    1.49 &    0.21\\
 SN 1996bo &    $-$12.05 &       1.45 &      0.10 & $-$13.61  &    1.64 &   0.10\\
\hline
\hline
\multicolumn{7}{l}{$^a$ age relative to $B$ maximum, days. } \\
\multicolumn{7}{l}{$^b$ mag relative to $B$ maximum, days. } \\
\multicolumn{7}{l}{$^c$ age relative to $V$ maximum, days. } \\
\multicolumn{7}{l}{$^d$ mag relative to $V$ maximum, days. } \\
\end{tabular}
\end{center}
\end{small}
\end{table}

\begin{table}[h]
\begin{small}
\begin{center}
\vspace{0.4cm}
\begin{tabular}{llll}
\multicolumn{4}{c}{Table 5: SN Ia Parameters} \\
\hline
\hline
SN Ia & $\Delta^a$ &  $\Delta m_{15}(B)$ & $s$ \\
\hline
SN 1996by & 0.25(0.05) & 1.37(0.06) &  0.85(0.02) \\
SN 1996bo & 0.21(0.05) & 1.22(0.06) &  0.93(0.02) \\
SN 1996bv & -0.32(0.07) & 0.94(0.08) & 1.14(0.03) \\
SN 1997bq & 0.14(0.05) & 1.23(0.05) &  0.89(0.02) \\
SN 1998dh & 0.14(0.05) & 1.23(0.05) & 0.94(0.02) \\
SN 1998ef & 0.06(0.05) & 1.29(0.05) & 0.92(0.01) \\
SN 1998bu & 0.02(0.05) & 1.15(0.05) & 0.96(0.01) \\
SN 1998aq & 0.10(0.05) & 1.12(0.05) & 0.94(0.01) \\
SN 1990N  & -0.33(0.05) & 1.03(0.06) & 1.02(0.02) \\
SN 1994D  & 0.39(0.05) & 1.40(0.05) & 0.82(0.01) \\
\hline
\hline
\multicolumn{4}{l}{$^a$ $\Delta=0$ when $M_v$=$-$19.34 (Jha et al. 1999).} \\
\end{tabular}
\end{center}
\end{small}
\end{table}

   The result, as seen in Figure 4, is illuminating.  Systematic
   differences in these prototypical SN Ia light curves are readily
visible.     Particularly noteworthy is the
apparent correlation between the rate of the early rise and the subsequent 
decline.  This phenomenon is
exemplified by 
the well-sampled SN 1990N; it appears to rise and decline
15\% to 20\% more slowly than SN 1994D.
The
   sense of this correlation has been extensively noted by other workers (Phillips
1993; Hamuy et al. 1995, 1996; Riess, Press, \& Kirshner 1995, 1996;
Riess et al. 1998b; Tripp 1997, 1998), except these data extend to much earlier ages than was previously
explored, leaving little chance that a discontinuity or break
from this rule occurs at earlier times.  
Because it has been previously demonstrated that the SN Ia decline
rate is correlated with SN Ia luminosity (Phillips
1993; Hamuy et al. 1995, 1996; Riess, Press, \& Kirshner 1995, 1996;
Riess et al. 1998b; Tripp 1997, 1998), we may also
conclude that brighter SNe Ia rise more slowly and explode
longer before maximum than do faint SNe Ia.   

  These conclusions are
only qualitative.  Now we proceed to a quantitative of exploration the early SN Ia light curve.
To accomplish this we
need a quantitative model of the risetime behavior and its variation. 

\subsection{Risetime Model}

To estimate the risetime we need a model 
of the rising light curve of any SN Ia 
which can be used to estimate the time of explosion.  A
very simple and sensible model has been presented by Goldhaber (1998) to fit the early phase of the SN Ia rise:
it is motivated by the theoretical representation of a SN Ia
shortly after explosion as an expanding fireball.  Through a passband
system on the Rayleigh-Jeans tail of the nearly thermal SED of a newborn SN Ia, the luminosity of
the expanding fireball will be
\bq L \propto v^2(t+t_r)^2T, \eq where $v$ is the photospheric velocity, $T$
is the temperature (or the model temperature for a dilute blackbody),  $t$ is time relative to maximum,
 and $t_r$ is the risetime (i.e., the time interval between the explosion and the maximum).  In the first few days after
explosion, the time since explosion, $t+t_r$, will increase by many factors.
This is in contrast to the {\it relative} stability of the temperature
(corroborated by the slow change in SN Ia colors) and the velocity
(corroborated by the slower fractional change in the wavelengths of absorption minima; Wells et al. 1994; Patat et al. 1996).  Hence at very early
times the luminosity should be roughly proportional to the square of the time since
explosion: \bq L=\alpha (t+t_r)^2, \eq  where $\alpha$ is the
``speed'' of the rise.  (For our empirical model, the speed of the
   rise is the change in luminosity over a fixed time interval.)  Although this model is motivated by 
theoretical arguments, it is empirically justified by the well-sampled early rise of SN 1990N (see Figure 4). 
This model assumes that at the explosion time the
luminosity of a SN Ia is zero.  In principle, the initial luminosity is
that of a white dwarf ($M_B$=10 to 15 mag), but at the observed speed of
the rise, the brightening from zero to a white dwarf luminosity
requires less than 1 second.  

\subsection{Inhomogeneity}

It is clear from Figure 4 that the risetime behavior of SNe Ia is
considerably inhomogeneous and we cannot expect any single set of
parameters, ($t_r$,$\alpha$), in equation (10) to suffice for our
sample.  Yet the rise is also likely related to the subsequent
shape or decline of the light curve.  

A general and economical hypothesis is that the rise parameters for an individual SN Ia are a linear function of the subsequent light-curve
shape \bq t_r=R X - t_0 \eq and \bq \alpha= S X + \alpha_0, \eq where
$X$ is any parameter which quantifies the post-rise light-curve shape and $R$ and $S$ are linear correlations coefficients between $X$ and the risetime and speed, respectively. 
 Substituting equations (11) and (12) into equation (10) yields a
custom rise model for any SN Ia whose individual risetime and speed depends
only on the light-curve shape parameter $X$: \bq
L=(SX+\alpha_0)(t+R X-t_0)^2. \eq  For a sample of SNe Ia there are four free
parameters: the ``fiducial'' risetime, $t_0$, the
fiducial speed, $\alpha_0$,  and the
linear correlation coefficients between $X$ and the risetime, $R$, and the speed, $S$.  

Two popular methods for
quantifying SN Ia light-curve shapes are the decline rate method
(Phillips 1993; Hamuy et al. 1996b) and the multicolor light-curve shape
method (MLCS: Riess, Press, \& Kirshner 1996; Riess et al. 1998b).
The former method employs the parameter $\Delta m_{15}(B)$, 
the decline in $B$ mag from peak to 15 days hence.  Typical SNe Ia
have $\Delta m_{15}(B)=1.1$ mag.  The MLCS method
correlates the differences between individual light-curve shapes and a
fiducial template shape with the differences between individual peak visual
luminosities and the fiducial peak luminosity, $\Delta \equiv
M_V-M_V(fiducial)$.  Jha et al. (1999) and Riess et al. (1998a) adopted 
$M_V(fiducial)$=$-$19.34 mag.
The values of the parameters $\Delta m_{15}(B)$ and $\Delta$
for our sample are given in Table 5.

It is sensible to define the fiducial SN Ia (i.e., $X=0$) to match the most commonly observed SN Ia, so
that the fiducial risetime parameters will characterize the mode of the population.
Because the luminosity calibration of typical SNe Ia is still subject to change, we
adopt the simple light-curve based criterion $\Delta m_{15}(B)=1.1$ mag to characterize the fiducial SN Ia.
For the purpose of comparing the two
light-curve methods we set $M_V(standard)$ (i.e., $\Delta=0$) to
be the luminosity of a SN Ia with $\Delta m_{15}(B)=1.1$ mag.  With this choice
both methods will be measuring the same fiducial risetime.  
Using Cepheid
calibrations of a set of nearby SNe Ia, Saha et al. (1999) find
$M_V=-19.45$ mag for a SN Ia with $\Delta m_{15}(B)=1.1$ mag.

 We tried both methods in our rise model by substituting
$X=\Delta$ and $X=(\Delta m_{15}(B)-1.1)$ into equation (13).
  Because the MLCS method derives $\Delta$ from a simultaneous
fit to all available light curves, we used $X=\Delta$ to model 
the rise data transformed to both $B$ and $V$.  In contrast our values of 
$\Delta m_{15}(B)$ were derived solely from the $B$ light curves
(using the parameter definition), so this parameter was only used to
model the $B$-band rise. 

  The most likely values for the free parameters in equation
(13) were found by minimizing the $\chi^2$ statistic, the standard measure of
the quality of the fit between model and data. 
We simultaneously fit all data earlier than 10 days before maximum.  After 
10 days before maximum our simple rise model fails to
   describe the light curve (see Figure 4).  
   As illustrated in Figure 5, the age by which 
we discontinue the fit does not impact the measured risetime until 
  approximately 8 days before maximum, when the value of
   $\chi^2_\nu$ also rises dramatically.  By this age it is not
surprising that the model would fail: the fractional time since explosion is no
longer changing rapidly compared to the velocity and temperature.  We have
conservatively chosen to fit only data at 10 days before
maximum and earlier to avoid biasing the fit by the inadequacies of the model.
In \S 4 we discuss the relevance of
the correlations of
the uncertainties in individual observations.  

All fits yielded values of $\chi^2$ within the expected statistical
range. The values of the parameters are given in Table 6.

  Although the speed
parameter, $\alpha$, is useful for modeling the rise, it has little
physical meaning.  The speed is an unknown function of the photospheric
velocity, temperature, and dilution factor.  The data do favor the existence
   of a correlation between the speed of the rise and the light-curve
   shape parameters in the sense that brighter and more slowly declining SNe
Ia are slower risers (i.e., smaller $\alpha$).  Unfortunately, the simplicity
of our model provides no basis to determine if the value or
variation of the speed has any physical significance.

Of greater physical interest
is the fiducial risetime and the correlation between the individual
risetime and the light-curve shape.  From these parameters we can determine 
the explosion time of a SN Ia or equivalently the age when observed.  To examine these parameters
we converted the $\chi^2$ into a probability density function (PDF; Lupton
1993) and marginalized the PDF over the speed parameters
$\alpha_0$ and $S$.  The resulting confidence intervals for $t_0$ and
$R$ are shown in Figure 6.  

Our statistical confidence that the SN Ia risetime is correlated
   with decline rate and the MLCS method's surrogate parameter for
   light-curve shape, visual peak luminosity, is very high (16$\sigma$
   and 9$\sigma$, respectively).

   Both methods concur that the risetime to $B$ maximum of a typical SN Ia (i.e., one
with $\Delta m_{15}(B)=1.1$ mag
or $M_V=-19.45$ mag) is 19.5$\pm$0.2 days.  The MLCS method gives a
fiducial risetime to $V$ maximum of 21.1$\pm0.2$ days or 1.6 days longer than the $B$-band risetime.
Errors in the peak luminosity
calibration of SNe Ia do not affect the risetime but rather the
appropriate luminosity for the fiducial risetime. 

 Using the MLCS
method we find that the brightness of a SN Ia is correlated with the
risetime in the sense that for every 0.10 mag 
increase in visual luminosity, SNe Ia 
require 0.80$\pm0.05$ and
0.77$\pm0.05$ days longer to
reach maximum in $B$ and $V$, respectively.  We emphasize that this
linear relation was determined over a modest range of luminosity
($-$19.6$ < M_V < $ $-$18.9) and should not be extrapolated beyond
this range.  Using the decline-rate parameter, we find that slower
declining SNe Ia require longer to reach $B$ maximum at the rate of
1.2$\pm0.1$ days for a 0.1 mag change in $\Delta m_{15}(B)$.  Again,
this relation should not be extended beyond the range over which it 
was derived, 1.39 $ > \Delta m_{15}(B) > $1.03.

  The difference between the $B$ and $V$ fiducial risetimes
from MLCS is 1.6 days. 

\subsection{Light Curve Normalization}

  An alternate method for measuring the risetime of SNe Ia has been
  proposed by Goldhaber (1998) based on the ``stretch'' method
  employed by Perlmutter et al. (1997).  The basis of this approach is
  to first normalize the inhomogeneous light-curve shapes and by assumption,
  the rise behavior, by dilating the
  observation times to fit the light curves to a fiducial template light curve.  
  This method is less general than our treatment in \S 3.2 because it
  {\it assumes} that the rise behavior is correlated with the
  rest of the light-curve shape in this simple way.  Examination of
  Figure 4 makes this assumption seem plausible;  however, the results
  of \S 3.2 indicate that it may be oversimplified.
  Although the timescale of the rise may be dilated in a similar
  fashion as the subsequent light curve, the speed of the individual rises also varies.  
  Nevertheless, the reward for employing this method is a reduction of
  the free parameters of the sample to only two: the risetime ($t_0$) and the speed
  ($\alpha_0$) of the fiducial template. 

  To find the fiducial risetime with this method we adopted the template from Leibundgut (1989) 
as the fiducial template.  Because this template light curve has $\Delta
  m_{15}(B)$=1.1 mag, the fiducial rise parameters should be directly comparable
to the values derived in \S 3.2 (assuming the template is otherwise characteristic of typical SNe Ia).  

   We have employed the stretch method of Perlmutter et al. (1997) by
   fitting each light curve to the fiducial template.  Three parameters
   and their uncertainties were simultaneously fit: the epoch of maximum, the
   peak magnitude, and the stretch factor, $s$.
   In all cases, {\it data 10 or
   more days
 before maximum were excluded
   from the fit to the light curves} so that the
   risetime behavior is normalized by the
   post-risetime data independent of the rise.  Further, because we do not know {\it a priori}
   the risetime behavior of the fiducial SN Ia, we cannot compare the risetime
   observations to a fiducial template to aid in the determination of
   the light-curve parameters.  In addition, we have not employed data
   more than 35 days after maximum since the stretch method fails to
   adequately describe the range of SN Ia behavior at these late
   times (Perlmutter et al. 1997).  The stretch parameters are listed
   in Table 5.

    By applying the fitted parameters
   to the entire light curve (now including observations earlier than
   10 days before maximum), each SN Ia light curve is normalized to
 a common fiducial template, as shown in Figure 7.  The reduction in photometric dispersion of the 
rise data is
   impressive.  Because the light-curve fits excluded the use of 
   data more than 10 days before maximum, the reduction in dispersion
   of these data attests to the correlation between the rise and the subsequent light curve shape.

\begin{table}[t]
\begin{small}
\begin{center}
\vspace{0.4cm}
\begin{tabular}{llllllll}
\multicolumn{8}{c}{Table 6: SN Ia Risetime Fits} \\
\hline
\hline
Method & Band & F.C. &  $\chi^2_{\nu}$ & $t_0$ & $R$ & $\alpha_0$ & $S$ \\
\hline
MLCS & $B$ & $M_V=-19.45$ & 1.0 & 19.42$\pm0.16$ & 8.0$\pm0.5$ &
0.078$\pm0.002$ & 0.040$\pm0.006$ \\
Decline rate & $B$ &$\Delta m_{15}(B)=1.1$  & 1.3 & 19.46$\pm0.23$ & 11.7$\pm1.3$ &
0.072$\pm0.003$ & 0.032$\pm0.011$ \\  
Stretch & $B$ & LT & 0.9 & 19.42$\pm0.19$ & --- &
0.071$\pm0.004$ & --- \\  
MLCS & $V$ & $M_V=-19.45$ & 1.4 & 21.02$\pm0.16$ & 7.7$\pm0.5$ &
0.074$\pm0.002$ & 0.020$\pm0.006$ \\  
\hline
\hline
\end{tabular}
\end{center}
\end{small}
\end{table}

A lower bound to the typical SN Ia risetime is immediately available from the
earliest observations.  The earliest SN Ia observation (normalized to the
Leibundgut template) is the observation by C. Faranda of SN 1998bu at
18.15 $\pm0.25$ days before and 4.63$\pm0.16$ mag below $B$ maximum.
  A coeval
observation (again as normalized to the Leibundgut template) of SN 1997bq was 
obtained by BAO at 18.15 $\pm0.35$ days
before and 4.0$\pm0.17$ mag below $B$ maximum.  From the slope of the
rise we see that at this age  the $B$ luminosity of a SN Ia is increasing 
5\% every hour.  From a weighted average of these two
earliest observations, we conclude that the risetime to $B$ maximum (for a SN Ia resembling the Leibundgut template) must be greater
than 18.15$\pm0.22$ days.  

   We fit equation (10) to the normalized rise data to find
   the fiducial risetime and speed of the template.  In accordance with the results in Figure 5, we 
only fit data earlier than 10 days before $B$ maximum.  
To fit the model to the data we converted the individual age uncertainties
of the data points to magnitude uncertainties using the local tangent
of the model (i.e., $\sigma_L=2\alpha\sigma_t$: Press et al. 1992).  The fit to the data is shown
   in the left panel of Figure 7.  

   The confidence
   intervals of the parameters ($t_r,\alpha$) are shown in the right panel of Figure 7.
    The values of the goodness-of-fit statistic,
$\chi^2_{\nu}$, listed in Table 6, show that the risetime behavior is
   well modeled by equation (10).
The result is that a SN Ia with a light-curve
shape similar to the Leibundgut template would explode $19.42\pm0.19$ days before  $B$ maximum (although the individual risetimes range from 16 to 21 days).  This
result is derived using the theoretical SN Ia pseudo-colors, but the
empirical SN Ia pseudo-colors give a consistent result with a risetime
   shorter by 0.18 days.  

   This result is in excellent agreement with the risetime value of
   19.5$\pm0.2$ days to $B$ maximum for typical SNe Ia derived in the
previous section.

\subsection{Non-detections}

   In addition to the $\sim$30  detections of SNe Ia earlier than 10
days before $B$ and $V$ maximum, we have 4 additional observations of the host
galaxies obtained near the time of explosion in which SN
light is not present.  For these observations, we  
determined the detection
limits empirically by adding artificial stars at a range of
brightnesses above and below the approximate detection level.
  Unfortunately, these upper limits provide 
negligible additional constraints on the rise behavior; the latest 
limit is one-half day after the fitted time of explosion when the
SN Ia (SN 1998dh) would be $\sim$ 7 mag below its peak in $B$ or $V$,
but the distance of this object results in a lower limit
which is only  $\sim$ 4 mag below peak.  To
detect a SN Ia at this age would require both an SN Ia as near as
the Virgo cluster and a telescope which can reach $\sim$ 19 mag in 
search mode (i.e., short exposures).  Although the upper limits
 presented here do not add to our constraints on the SN Ia rise behavior, in
the future such observations may yield useful results; on good nights, for example, the LOSS 
images yield detections at $\sim$ 19 mag.

\section{Biases and Systematic Errors}

   In this section we consider sources of systematic errors which
   could bias our measurement of the SN Ia
   rise behavior.  Because the stretch method gives consistent results with
   the method in \S 3.2 but is less computationally intensive to
   implement,  we use this method in the following tests of systematic errors. 

\subsection{Selection Bias}

   A bias in our risetime measurement
can result from an intrinsic dispersion in
the risetime of SNe Ia with the same post-rise light-curve
shape.  Although we have already normalized the rise behaviors for the
variations in the one-parameter family of
light-curve shapes, some dispersion in the risetime might exist for
SNe with the same subsequent light-curve shape.  Given a variation of risetimes,
 SNe which spend more time above a detection limit before maximum will preferentially
be discovered and included in our sample.  This would bias the average towards
``slow risers'' (i.e., long risetimes.)

As seen in Table 6, the
values of $\chi^2_\nu$ for the fits to the rising data are all within their expected statistical
range indicating that there is little or no
intrinsic dispersion of the risetime after normalizing the light-curve shapes.   
If no intrinsic dispersion is evident at the resolution of
the data, then the bias on the measured rise behavior can be neglected.
Nevertheless, we can use an additional 
criterion to select a subsample of SN Ia data from
which we can infer the unbiased risetime.  This criterion only includes SNe which were discovered
later than 10 days before maximum.  This condition guarantees that the risetime
of any SN Ia did not influence its discovery and hence its
inclusion into our unbiased subsample.
The SNe Ia in our full sample which fail this criterion are SN 1990N, SN 1994D, and SN 1998aq\footnote[2]{However, an argument can be made that SNe Ia in the three major nearby clusters would be
 discovered at early phases regardless of their risetime.
  SN 1990N and SN 1994D both occured in
 the Virgo cluster, the former found by amateur astronomers and the latter by a
 professional search.  The Virgo cluster is a major target of amateur
 and professional supernova searchers and is monitored with high frequency.
 In the Spring of 1989 a group of amateur astronomers initiated the
 ``Virgo Project'' to systematically search for SNe in Virgo (Starburst
 Newsletter 20 from VSNET).  This high frequency of monitoring should
 result in the discovery of nearly all typical SNe Ia independent of their risetime.
 The same conclusion
 may be appropriate for SN 1998aq which occured in the Ursa Major
 cluster.  Nevertheless, it is wise to temporarily discard these objects to be
 sure that our subsample will yield an unbiased estimate of the risetime.}.

 Even SNe Ia which were discovered later than 10 days before
 maximum, yet were detected in pre-discovery images, must
 be scrutinized.   To insure an unbiased
 measurement of the risetime it is necessary to measure all
 pre-discovery observations of the host galaxy regardless
 of whether SN light is clearly present.  One of our earliest SN detections, that
 provided by C. Faranda of SN 1998bu, was posted on the amateur SN
 search network, VSNET (and brought to our attention by
 P. Meikle).
  Yet it may be that this report exists {\it only because the SN was evident} in the
 observation.  Had the SN not been present in the observation, the
 null detection might not have been reported and this object would not
 have been included in our sample.  The selection of an early
 observation due to the presence of SN light can lead to a
 bias towards slow and long risers (for a given light-curve shape).  Therefore,
 we conservatively discarded SN 1998bu when accumulating our unbiased sample.

 What remains in our unbiased sample are only the SNe discovered by LOSS and
 BAO: SN 1996bo, SN 1996bv, SN 1996by, SN 1997bq, SN 1998dh, and SN
 1998ef.  All of these objects were discovered later than 10 days
 after $B$ maximum and all pre-discovery images of the
 host galaxy were available for measurement without regard to the
 presence or absence of SN light.   From such data we can insure that
 our measurement of the rise behavior is unbiased.

   The lower-precision results from the unbiased data set are highly consistent with the
 results from the complete sample.  We find a risetime to $B$ maximum 19.62$\pm0.40$
 days, in good agreement with the value of 19.42$\pm0.19$ from the full
 set of SNe Ia data.  We conclude that no bias towards longer risetimes is evident 
from the full sample.

\subsection{The Color of a Newborn SN Ia}

 Systematic errors in the unfiltered SN Ia transformations could
 arise from a significant failure of our assumption that
   the $B-V$ colors do not evolve between our earliest observations at
   16 to 18 days before maximum ($\sim$ 2 days after explosion) and 
14 days before maximum by which time SN Ia $B-V$ colors have been
   observed (Leibundgut et al. 1991a; Richmond et al. 1995; Riess et
   al. 1999a; Hamuy et al. 1996a).  
   We can remove the influence of this assumption in numerous ways.

  One method is to transform some
   of the unfiltered observations to the $R$ passband.  
   Over the range $0.0 < B-V < 1.5$ the coefficient of the linear
   transformation is $C_{RW}=0.0\pm0.05$ mag for E. Thouvenot's CCD.
  For any value of $B-V$ of a young SN Ia 
 within this range, the transformation to $R$ is independent of
   $B-V$.  We inferred the explosion time from the data of SN 1990N
 transformed to $R$ compared and measured
   the time interval to $B$ maximum.  
   (This method assumes that the time of explosion can be inferred equivalently 
from observations in different passbands).
The resulting risetime differed by
   less than 0.05 days with the value inferred from the
   transformations to $B$.  

   We also refit the $B$-band risetime using only data between 10 and 14 days before $B$ maximum.  The
   earliest observations of SN Ia colors are static over this range in age (Leibundgut et al. 1991a; Riess et al. 1999a, 2000; Richmond et al. 1995; Hamuy et al. 1996a; 
Li et al. 1999).
   Normalized to the SCP $B$ template, data between 10 and 14 days
   before maximum gave a risetime of 19.00$\pm$0.52 days, in reasonable
   agreement with the full data set result of 19.42$\pm$0.18 days.    
   
   Finally, an unexpected change in the young SN Ia SED or $B-V$
   color would have a much larger impact on data transformed to $B$
   than to $V$.  The reason is that the $V$-band is a better match to the unfiltered CCDs and so the transformation depends
less on the form of the SN SED.  A systematic difference between the
   explosion time inferred from transformations to $B$ and $V$ would indicate an unaccounted for change in the SN Ia SED.  However, the measured difference between the risetimes to $B$ and $V$ maxima
   from MLCS of 1.6 days is in excellent agreement with the average
   difference in the times of $B$ and $V$ maxima of 1.64 days.
   Therefore, the difference in the explosion time inferred from the
   $B$ and $V$ transformations is less than 0.1 days and shows no
   evidence of a significant change in the early SN Ia SED.

\subsection{Other}

 Another source of potential error arises from the correlation of
 errors among individual risetime observations.  Specifically,
 all individual measurements of the same supernova would be expected
 to share some sources of error.  These include the
 determination of the stellar and supernova pseudo-colors and the
 light-curve fit parameters.   While it is possible to
 account for these correlations, an examination
 of our observations suggests a straightforward way to minimize the
 effects of error correlations.  Each SN Ia in
 our set has only one or two detections at early times, with one
 exception: SN 1990N has a
 total of 12 observations at 10 distinct ages.  By removing SN 1990N
 from our set we can remove the dominant source of covariance
 between observations in our sample.  This decreased the
 risetime by 0.3 days.  

\section{Discussion}
   
    Using a set of the earliest  detections of low-redshift SNe
   Ia, we have made a precise measurement of the risetime of
   SNe Ia. The low-redshift observations indicate that the rate
 at which SNe Ia rise
   and decline is highly correlated. 

The risetime to $B$ maximum for a typical SN Ia is 19.5$\pm
   0.2$ days, where ``typical'' is defined as having a peak of $M_V$=$-$19.45 mag,
$\Delta m_{15}(B)=1.1$ mag, or a $B$ light curve which resembles the
Leibundgut (1989) template.  Brighter and slower declining SNe Ia have
a longer risetime at the rate of 0.80$\pm0.05$ days per 0.10 mag in
peak luminosity and 1.17$\pm0.13$ days per 0.10 mag in $\Delta
m_{15}(B)$.  Most SNe Ia (i.e., with $-19.10 > M_V > -19.65 $ and $0.95 \leq
   \Delta m_{15}(B) \leq 1.40$) range in their
   risetime to $B$ maximum by 16 to 21 days depending on their
   specific light-curve shape or width.  

   We have not considered SNe Ia which are beyond this range 
for the reasons discussed in \S 2.1.  However, we have
   verified that the conclusions presented here do extend to highly
   subluminous SNe Ia using early observations of SN 1998de
 ($\Delta m_{15}(B)=1.96$ mag), a
   veritable twin to SN 1991bg (Modjaz et al. 2000).

   Our observations are not highly consistent with less precise
   measurements of the SN Ia rise behavior by Vacca \& Leibundgut (1996).
  They find the
   risetime to $B$ maximum for SN 1994D ($\Delta m_{15}(B)=1.32$ mag) 
to be 17.6$\pm0.5$ days as compared to 15.5$\pm0.5$ days found by
   fitting equation (10) directly to the $B$ data of SN 1994D.  This
   difference is not surprising since Vacca \& Leibundgut adopted a
   very different empirical model (a Gaussian multiplied by an
   exponential) to fit the rising light curve.  The relatively late
   start of the light curve of SN 1994D could result in large
   changes in the risetime for different methods used to extrapolate
   to the time of explosion.  An individual $B$ risetime measurement
   of 21.4$\pm$0.3 days for SN 1990N is somewhat larger than a
   ``loose'' upper limit of 20 days estimated by Leibundgut et al. (1991a) using an
   expanding photosphere approach.
   Our characteristic SN Ia risetime to $B$
   maximum of 19.5 days is 
   significantly longer than the first indications by Barbon, Ciatti,
   \& Rosino (1973) of 15$\pm2$ days.  Interestingly, our results are
in surprisingly good agreement with those of Pskovskii (1984) who found that typical SNe Ia have a risetime of 19 to 20 days and that an 0.1 mag increase in
the peak blue luminosity is accompanied by an $\sim$0.4 day increase in the risetime.   

   As noted by Vacca \& Leibundgut (1996), we find that most theoretical
   models of SNe Ia predict significantly shorter risetimes than we find.
  A series of model calculations by H\"{o}flich \& Khokhlov (1995)
   generally give risetimes to visual maximum of 9 to 16 days with an
   average value of about 14 days for single white dwarf progenitors.
   These models encompass a variety of explosion scenarios:
   detonations, deflagrations, delayed detonations, pulsations, and
   helium detonations.   The same difference is seen in
   model calculations by Pinto \& Eastman (1999).  

   Within a family of
   models, the predicted correlation between luminosity and risetime
   is opposite to what is observed.  Leibundgut \& Pinto (1992)
   calculated a series of classical deflagration as well as delayed
   detonation models and found that ``All models show the trend of
   decreasing luminosity for longer rise times.''  The same difference
   with the observations is true of the models by H\"{o}flich \& Khokhlov
   (1995).  

The double degenerate models of H\"{o}flich \&
   Khokhlov (1995) (modeled as a detonation within a thick envelope
   formed by a tidally disrupted companion) 
  can adequately reproduce the observed risetimes.  These models yield
   risetimes to $V$ maximum of 19.5 to 21.8 days, in good agreement with
   our fiducial $V$ risetime of 21 days.  This consistency could be
   seen as a reason to favor this candidate as the progenitor system of SNe Ia.
  However,
   the double degenerate models with the longest risetimes develop into the 
dimmest SNe, a trend in poor accordance with our findings.

  If these models are otherwise
   accurate, we concur with the conclusion made by Vacca \& Leibundgut
   (1996) that the model atmospheric opacity has been significantly
   underestimated.  Past work suggests that deficient resonance line
   lists may be the culprit.  By increasing the number of resonance
   lines from 500 to 100,000, the risetime for models by Harkness
   (1991) increased by 8 days.  More atomic data is needed to
   determine whether the discrepancy between models and observations
   can be resolved.  Ideally the risetime
   predictions of different models could be used to discriminate between proposed
   explosion mechanisms and progenitor types.  Unfortunately, 
   deficiencies of the current suite of
   models makes it difficult to access which model characteristics
   are favored at the present time.

   Recently, H\"{o}flich, Wheeler, \& Thielemann (1998) have found that
   reducing the carbon-to-oxygen ratio of the white dwarf progenitor 
from the previously assumed value of unity increases the risetime to
   values which appear to be a better match to the measurements reported here.
   These models have the advantage that SNe Ia with longer risetimes
   reach brighter peak luminosities, in agreement with our findings reveal. However, these models do not appear to match the well-known trend between peak luminosity and decline rate.  

   Assuming that the risetime to $B$ maximum is a good approximation
   of the bolometric risetime as found by Vacca \& Leibundgut (1996),
   we can use theoretical scaling laws from Nugent et al. (1995) to
   assess the peak absolute luminosity of SNe Ia.  Using the risetime
   to measure the rate of energy deposition from the radioactive decay
   of $^{56}$Ni and $^{56}$Co with a SN Ia bolometric correction to $B$
   band of 0.2 mag (H\"{o}flich \& Khokhlov 1996) we find
   $M_B$=$-$19.4$\pm0.3$ mag.  An alternate method using the risetime to
   measure the size of the dilute expanding photosphere at maximum
   gives $M_B$=$-$19.8$\pm0.3$ mag.  The two methods are in rough
   agreement with each other, as well as with current Cepheid based
   calibrations of the SN Ia peak luminosity of $M_B$=$-$19.45$\pm0.10$ from
   Saha et al. (1999). However, the qualitative differences between
   the observed and theoretically predicted risetimes cast
   doubt on the integrity of even semi-theoretical routes to the
   luminosity calibration of SNe Ia.

   We emphasize that an accurate determination of the risetime is not
   the limiting challenge for a theoretical determination of the
   fiducial peak SN Ia luminosity.  Within a single parameterized
   model, Leibundgut \& Pinto (1992) find that a change in the risetime by 1
   day results in a change in peak luminosity by 3\% (again, in the
   opposite direction to the observations reported here).  Yet, for a
   given risetime, they find that different models support a 50\% range
   in peak luminosity.   

   In the future, theoretical models need to be sharpened and reworked
   to match the increasingly precise observational constraints on the
   behavior of SNe Ia.  We hope that the characterization of the
   fiducial value and variation of the SN Ia risetime presented here
   can serve as a guide in the quest to develop a satisfactory
   theoretical understanding of SNe Ia.

      \bigskip
\bigskip 
 
We are indebted to Eric Thouvenot, Peter Meikle, and Chuck Faranda for 
providing the CCD images of young SNe Ia.
We wish to thank Ed Moran, Peter Nugent, Gerson Goldhaber, Saul Perlmutter, Don
Groom, Robert Kirshner, Peter Garnavich, Saurabh Jha, and
Doug Leonard for helpful discussions.  The work at U.C. Berkeley was supported by the Miller Institute for Basic Research
in Science, by NSF grant AST-9417213, and by grant GO-7505 from
the Space Telescope Science Institute, which is operated by the
Association of Universities for Research in Astronomy, Inc., under
NASA contract NAS5-26555.  We are also grateful to the Sylvia and
 Jim Katzman Foundation, Autoscope Corporation, the National Science Foundation, Photometrics, Ltd., Sun Microsystems Inc., and the University of California for assistance with the construction and suuport of KAIT at Lick Observatory.

\vfill \eject
 
\centerline {\bf References}
\vskip 12 pt

\refitem Armstrong, M., et al., 1998, IAUC 6497

\refitem Arnett, W. D., 1982, ApJ, 253, 785

\refitem Arnett, W.D.,Branch, D., \& Wheeler, J.C.,1985, Nature 314,337

\refitem Barbon, R., Ciatti, F., Rosino, S., \& Rafanelli, P., 1982,
A\&A, 116, 43

\refitem Barbon, R.,Ciatti,F., \& Rosino, L., 1973, A\&A 25,65

\refitem Branch, D., Fisher, A., \& Nugent, P., 1993, AJ, 106, 2383

\refitem Cardelli, J.A., Clayton, G. C., \& Mathis, J. S., 1989, ApJ,
345, 245

\refitem Filippenko, A.V. 1997, ARA\&A, 35, 309

\refitem Goldhaber, G., 1998, B.A.A.S, 193, 4713

\refitem Goldhaber, G., et al., 1999, in preparation

\refitem Gunn, J. E., \& Stryker, L. L., 1983, ApJS, 52, 121 

\refitem Hamuy, M., et al., 1996a, AJ, 112, 2408

\refitem Hamuy, M., Phillips, M. M., Suntzeff, N. B., Schommer, R. A., 
Maza, J., \& Avil\'es, R. 1996b,
AJ, 112, 2398

\refitem Hamuy, M., Phillips, M. M., Maza, J., Suntzeff, N. B., Schommer, 
R. A., \& Avil\'es, R. 1995, AJ, 109, 1

\refitem Hamuy, M., et al., 1991, AJ, 102, 208

\refitem Harkness, R. P., 1991, in SN 1987A and Other Supernovae, ed. I. J. Danziger \& K. Kjar (Garching:ESO), p. X

\refitem Ho, W., et al., 1999, PASP, submitted

\refitem H\"{o}flich, P.,Khokhlov, A., \& M\"{u}ller, E., 1992,
A\&A, 259, 549

\refitem H\"{o}flich, P., Wheeler, J. C., \& Thielemann, F. K., 1998,
ApJ, 495, 617

\refitem H\"{o}flich, P., \& Khovkhlov, A., 1996, 457, 500

\refitem Jha, S., et al., 1999, ApJSS, in press

\refitem Kim, A., Goobar, A., \& Perlmutter, S. 1996, PASP, 108, 190

\refitem Landolt, A. U. 1992, AJ, 104, 340

\refitem Leibundgut, B., et al., 1991a, ApJ, 371, L23

\refitem Leibundgut, B., Tammann, G. A., Cadonau, R., \& Cerrito, D.,
1991b, A\&AS, 89, 537

\refitem Leibundgut, B. 1989, PhD Thesis, University of Basel

\refitem Leibundgut, B., \& Pinto, P. A., 1992, ApJ, 401, 49

\refitem Li, W. D., et al., 1998, private communication 

\refitem Li, W., et al., 1996, IAUC 6379

\refitem Li, W., et al., 1999, AJ, in press

\refitem Lira, P., et al., 1998, AJ, 115, 234 

\refitem Lupton, R., 1993, ``Statistics in Theory and Practice,''
Princeton University Press, Princeton, New Jersey

\refitem Marvin, H., \& Perlmutter, S., 1989, IAUC 4727

\refitem Maury, A., et al., 1990, IAUC 5039

\refitem Meikle, P., \& Hernandez, M., 1999, Journal of the Italian Astronomical Society, in press

\refitem Modjaz, M., et al., 2000, in preparation

\refitem Nugent, P., Branch, D., Baron, E., Fisher, A., \& Vaughan,
T., 1995, PRL, 75, 394 (Erratum: 75, 1874)

\refitem Nugent, P., 1998, B.A.A.S, 193, 4712

\refitem Patat, F., et al., 1996, MNRAS, 278, 111

\refitem Perlmutter, S., et al., 1997, ApJ, 483, 565

\refitem Phillips, M. M. 1993, ApJ, L105, 413

\refitem Pinto, P. A., \& Eastman, R., 1999, ApJ, submitted

\refitem Press, W.H., Teukolsky, S.A., Vetterling, W.T., and Flannery, B.P. 1992, Numerical Recipes, 2ed (Cambridge University Press)
 
\refitem Richmond, M. W., Treffers, R. R., \& Filippenko, A. V., 1993, PASP, 105, 1164
\refitem Richmond, M. W., et al., 1994, IAUC 6105

\refitem Richmond, M. W., et al., 1995, AJ, 109, 2121

\refitem Riess, A. G., Press W.H., \& Kirshner, R.P., 1995, ApJ, 438 L17 

\refitem Riess, A. G., Press, W.H., \& Kirshner,  R.P. 1996, ApJ, 473,
88 

\refitem Riess, A. G., Nugent, P., Filippenko, A. V., Kirshner, R. P., \& Perlmutter, S.,
1998a, ApJ, 504, 935

\refitem Riess, A. G., et al., 1998b, AJ, 116, 1009

\refitem Riess, A. G., et al., 1999a, AJ, 117, 707

\refitem Riess, A. G., et al., 1999b, ApJ, submitted 

\refitem Riess, A. G., et al., 2000, in preparation

\refitem Saha, A., et al., 1999, ApJ, in press

\refitem Schmidt, B. P., et al. 1998, ApJ, 507, 46

\refitem Stone, R. P. S., \& Baldwin, J. A., 1983, MNRAS, 204, 347 

\refitem Treffers, R. R., et al., 1997, IAUC 6627

\refitem Tripp, R., 1997, A\&A, 325, 871

\refitem Tripp, R., 1998, A\&A, 331, 815

\refitem Vacca, W. D., \& Leibundgut, B., 1996, ApJ, 471, L37

\refitem Wells, L.A., et al., 1994, AJ, 108, 2233

\vfill \eject

{\bf FIGURE CAPTIONS:}

{\bf Fig 1.}-Wavelength response functions.  For each observer or observatory
 listed in Table 1, we show the CCD manufacturer's stated response
function multiplied by the atmospheric transmission including effects of glass and 
mirror transmission.  Overplotted for comparison are the Johnson/Cousins $B$ 
and $V$ response functions.  Also shown is the spectrum of SN 1990N at 14 days 
before $B$-band maximum light.     By accounting for the differences between the SN Ia SED observed by a 
standard passband and an unfiltered CCD it is possible to transform an unfiltered CCD
observation of an SN Ia to a standard passband.

{\bf Fig 2.}-For a set of 175 spectrophotometric stars given by Gunn \& Stryker (1983) we have calculated synthetic unfiltered CCD magnitudes for the BAO response function of
Figure 1.  A comparison of the synthetic $B-V$ colors and the difference between the $V$ and unfiltered magnitudes (open symbols) shows a well-defined correlation.   A similar relation is seen (filled symbols) by comparing the observed photometry of local standard stars in the fields of SNe.  The slope of this correlation defines a transformation between
unfiltered and standard passband magnitudes as a function of $B-V$ color.

{\bf Fig 3.}-A comparison of $B$ and $V$ SN Ia magnitudes observed through standard passbands and transformed to standard passbands using unfiltered CCD observations
and the theoretical transformation method.  For both SN 1990N and SN 1997bq the 
agreement between the standard and transformed magnitudes is excellent, with a 
relative dispersion of 0.03 and 0.04 mag in $V$ and $B$, respectively.

{\bf Fig 4.}-(a) Observed $V$-band light curves for 10 SNe Ia
including  risetime observations transformed from unfiltered CCD
measurements.  For each SN Ia, the time and magnitude of the peak has
been fit and subtracted from the data.  Significant inhomogeneity is
apparent including a strong correlation between the rate of rise and
decline to maximum.  (b) Same as for (a) except using observations and
transformations to the $B$ passband.  Additionally, an empirical model,
 equation (10),
 is shown as a fit to the data for SN 1990N.

{\bf Fig 5.}- $\chi^2_\nu$ and risetime versus the age (relative to $B$ maximum) by
which the fit to the rise data is terminated.  The fit remains reliable up to $\sim$8
days before $B$ maximum as evidenced by the acceptable values of $\chi^2_\nu$ and the relative 
stability of the risetime.  We conservatively use 10 days before maximum as the
age by which we terminate the fit.

{\bf Fig 6.}- Confidence intervals on the risetime parameters.  The upper panel shows
the likely values of the fiducial (i.e., peak $M_V$=$-$19.45 mag) 
risetimes to $B$ and $V$ maxima and the correlation between
the individual SN Ia $B$ and $V$ risetimes and the 
peak visual luminosity inferred using the MLCS method.
The lower panel shows the likely value of the fiducial (i.e., $\Delta m_{15}(B)$=1.1 mag)
  $B$ risetime and the relation between
the individual $B$ risetime and the decline rate parameter, $\Delta m_{15}(B)$.  

{\bf Fig 7.}-The early rise $B$ data normalized by the stretch method and the inferred risetime
parameters.  The observation times of the individual SNe Ia are dilated to provide the
best fit of the post-rise (i.e. after 10 days before maximum) data (diamonds)
 to the Leibundgut (1989) fiducial template.  Fitting equation (10) to the rise data (filled circles) 
yields the likely values for the speed of the rise and the risetime to $B$ maximum.

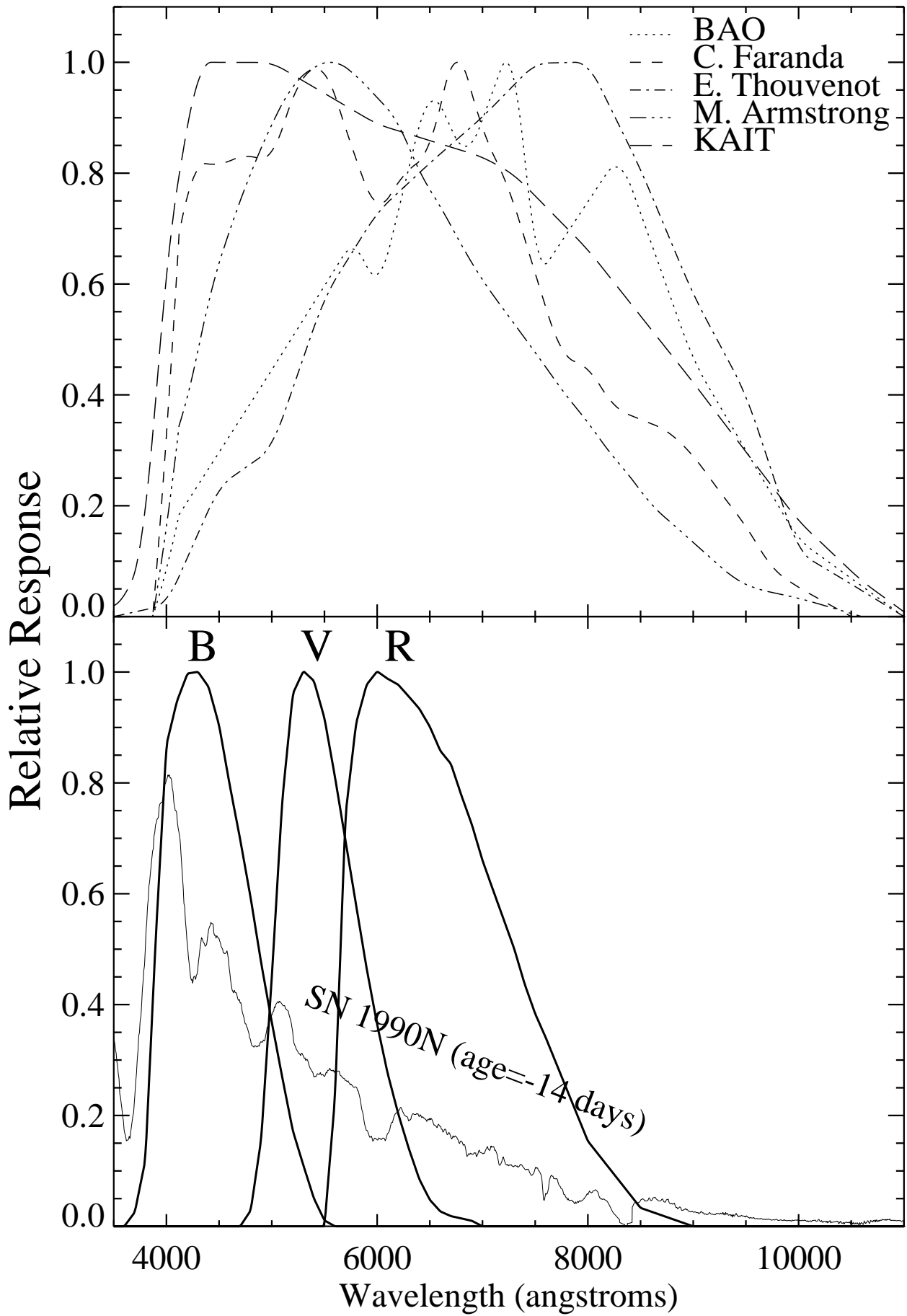
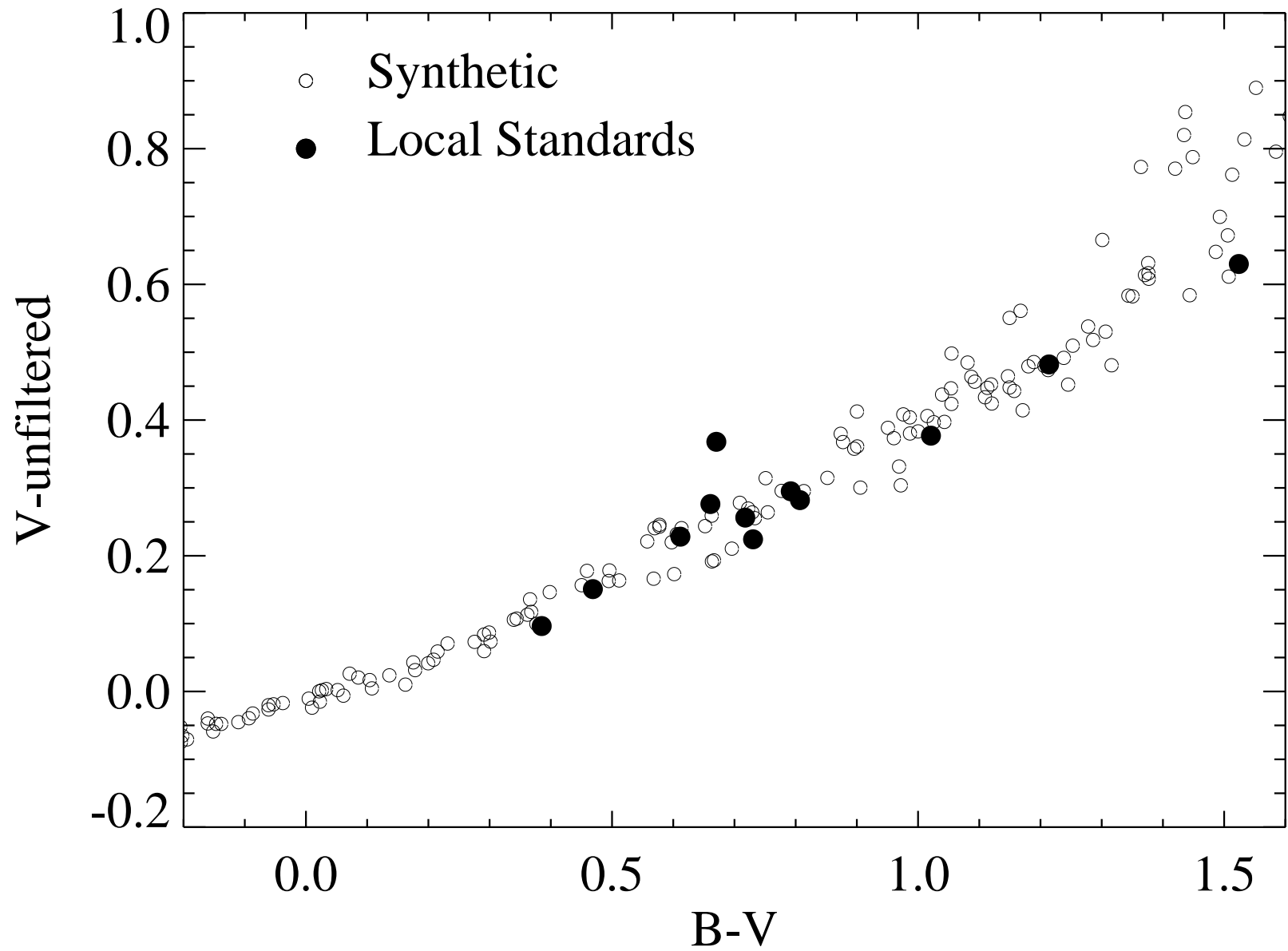
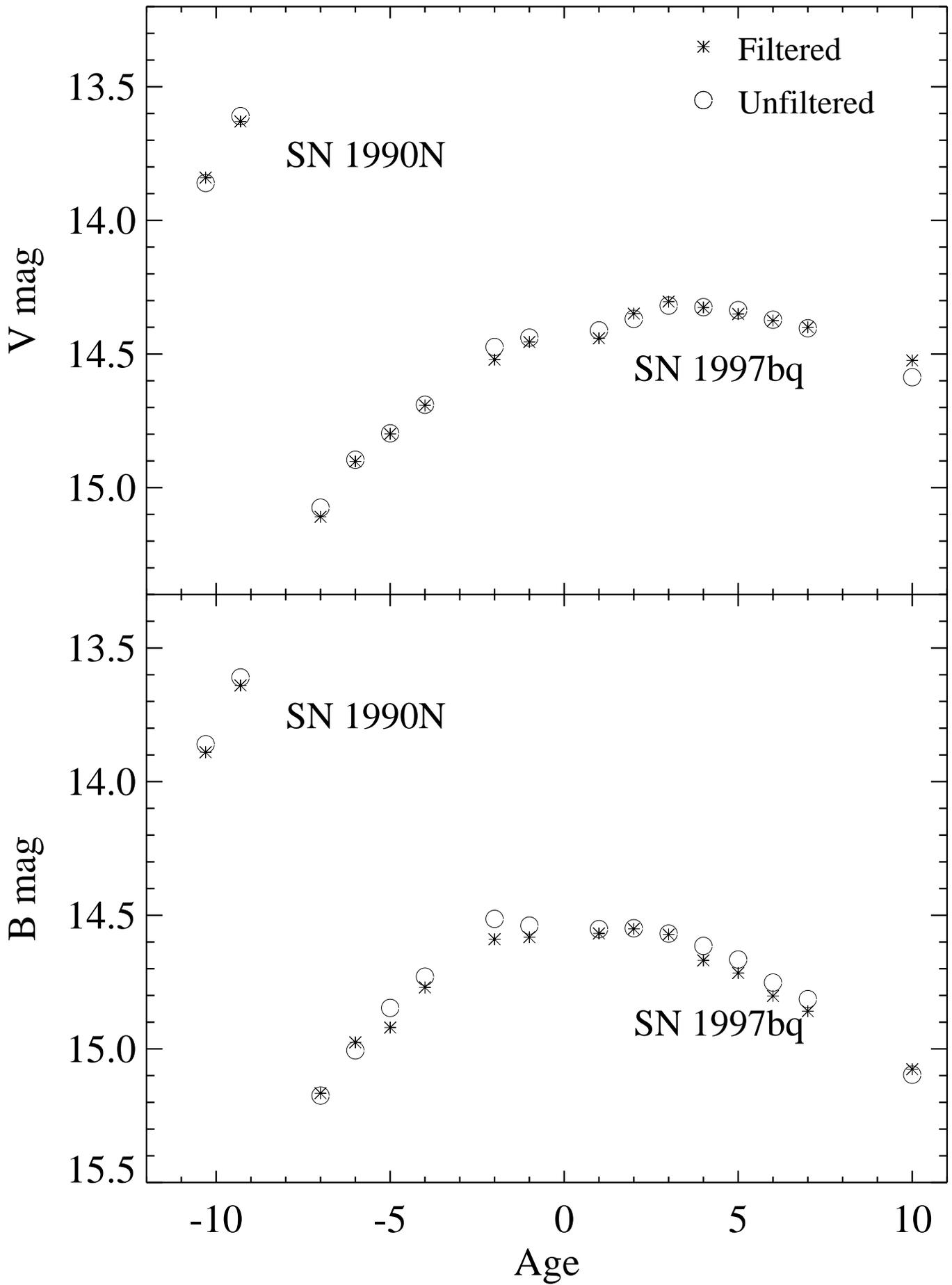
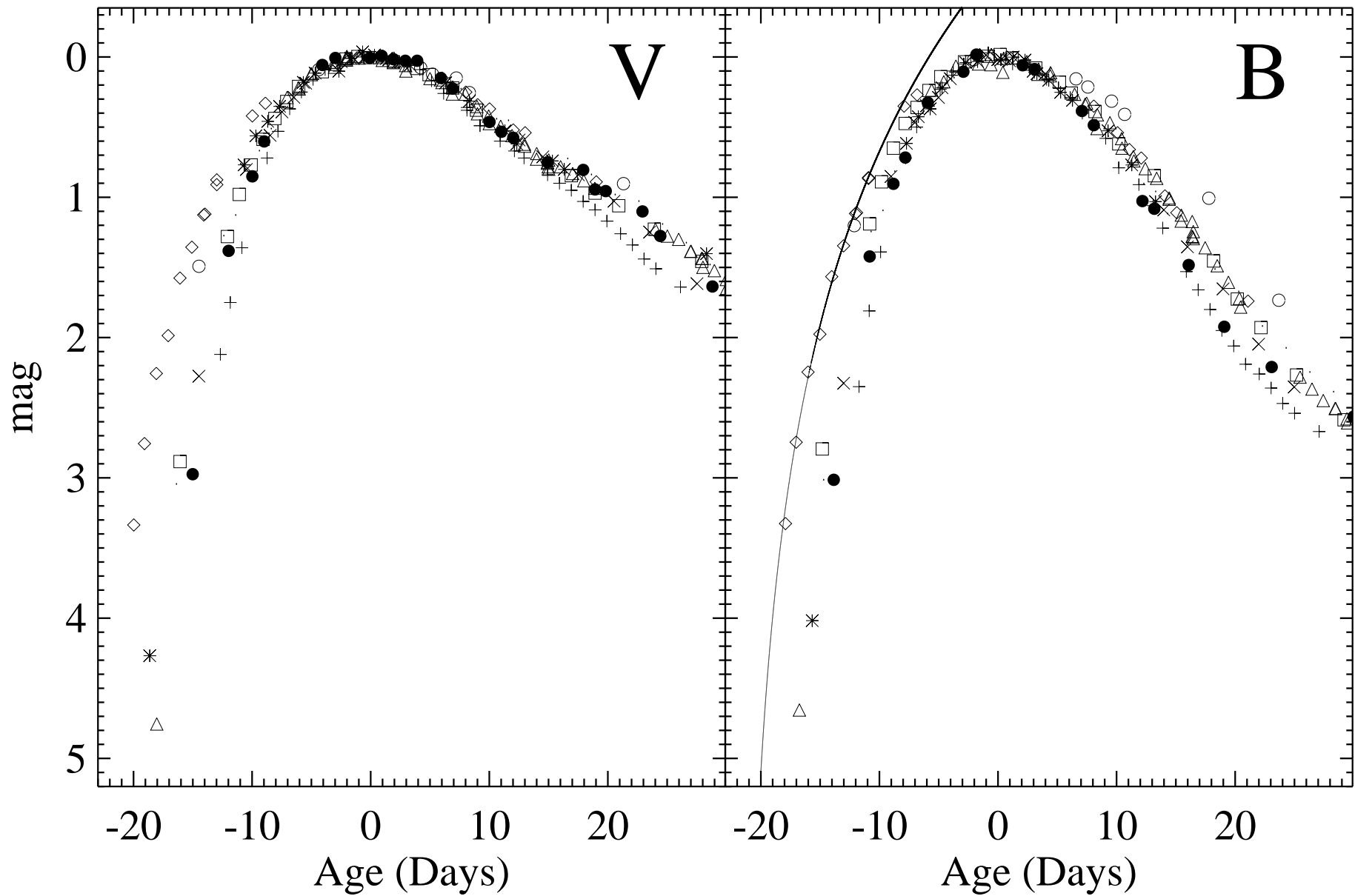
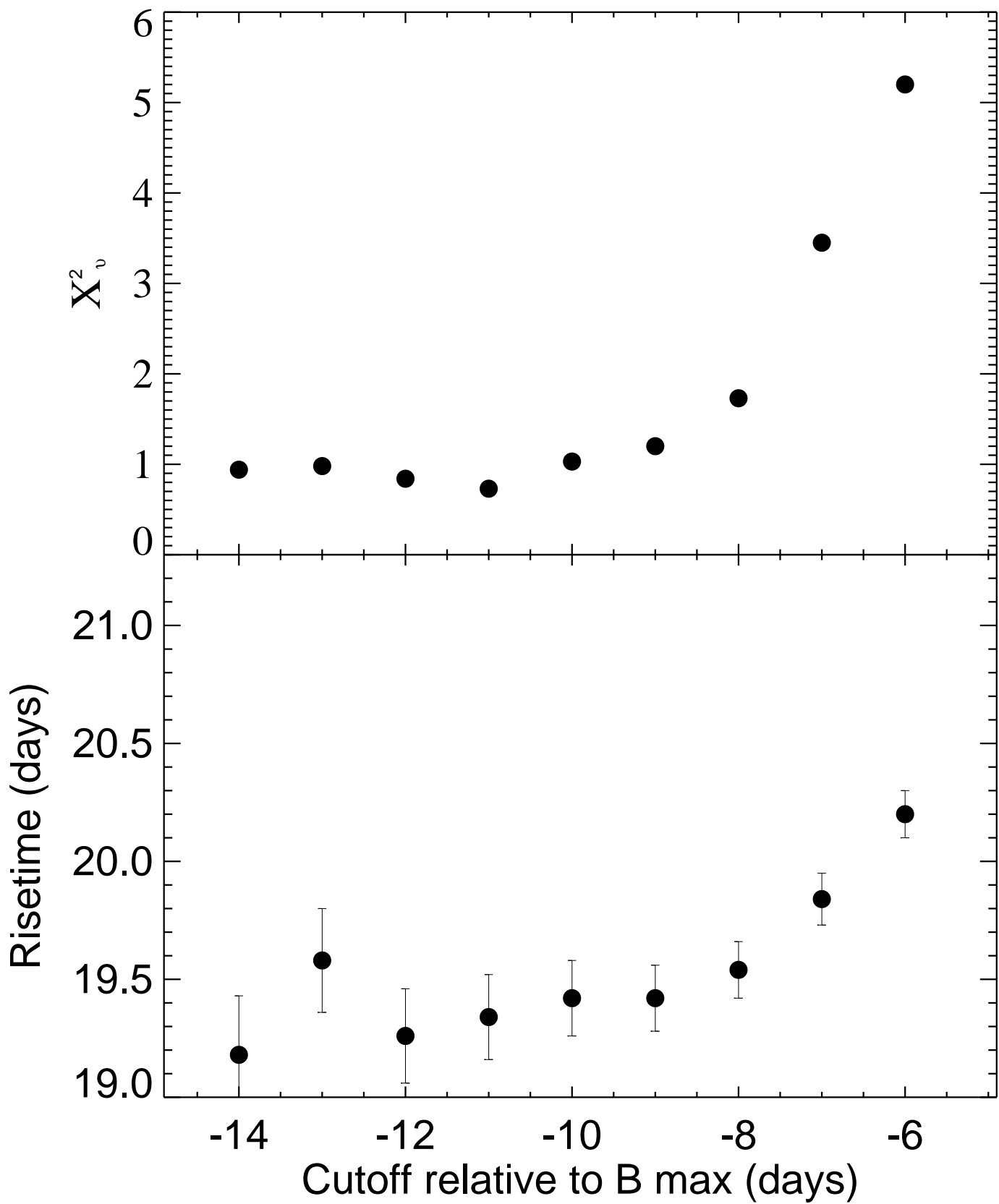
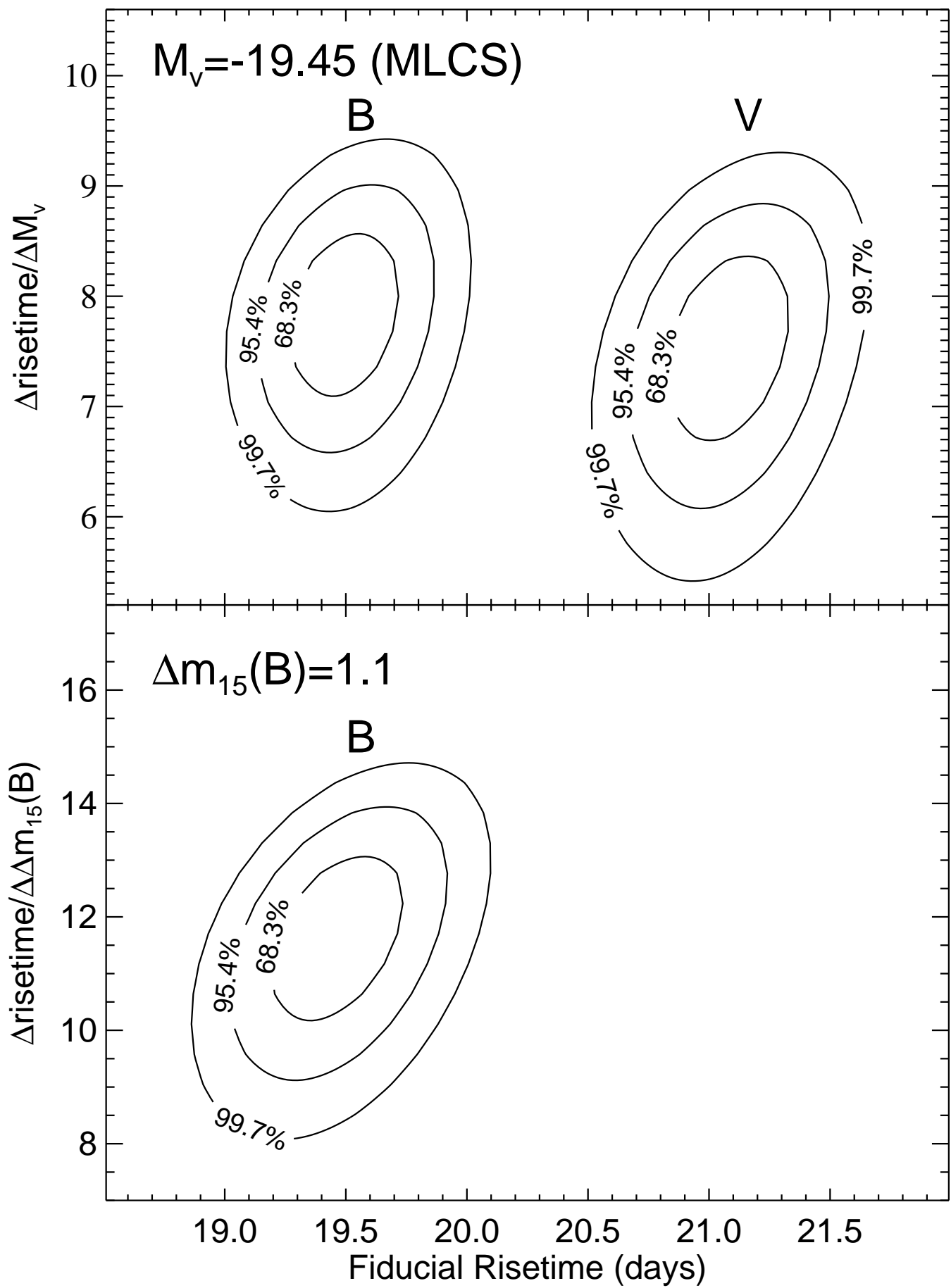
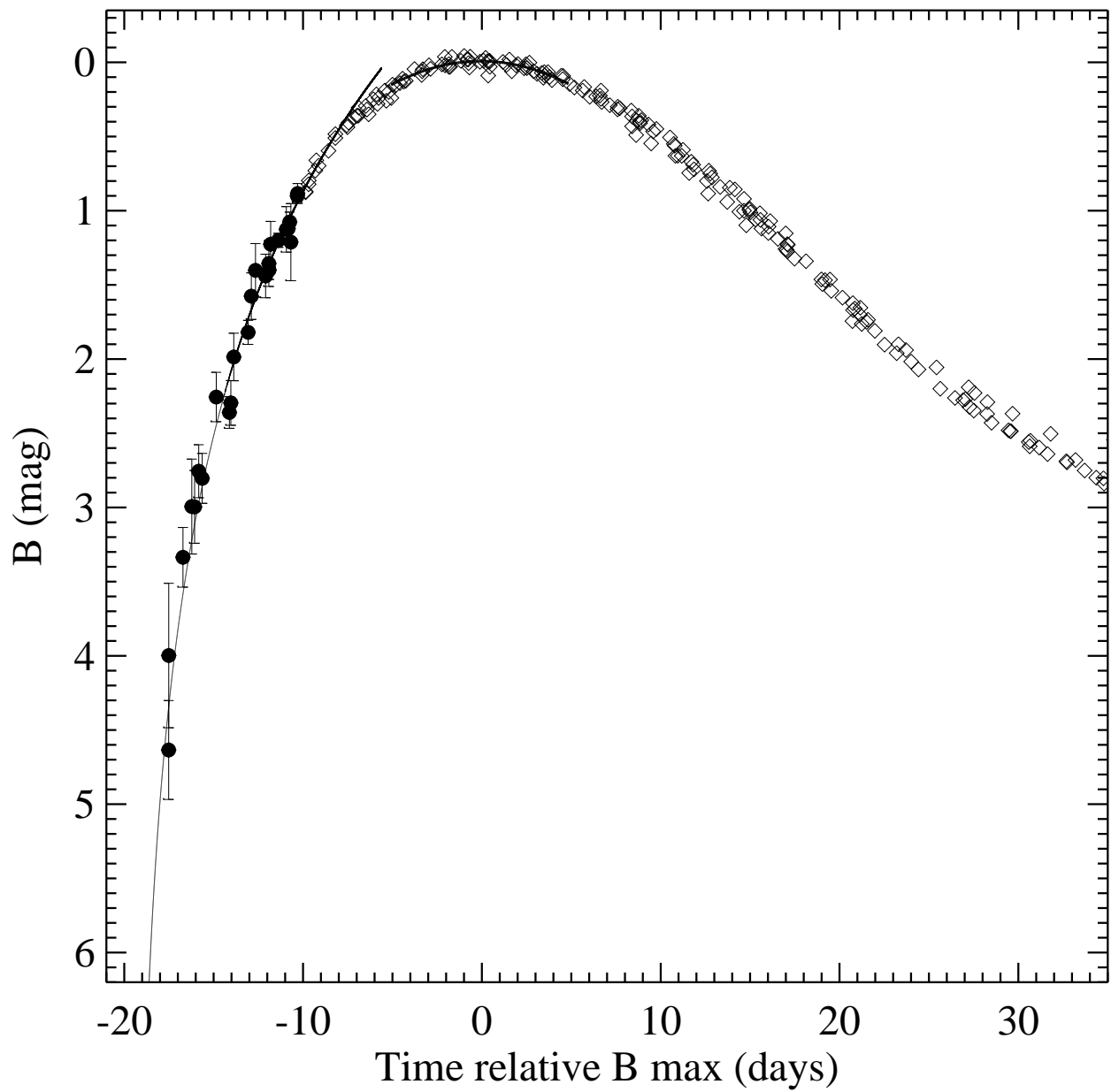
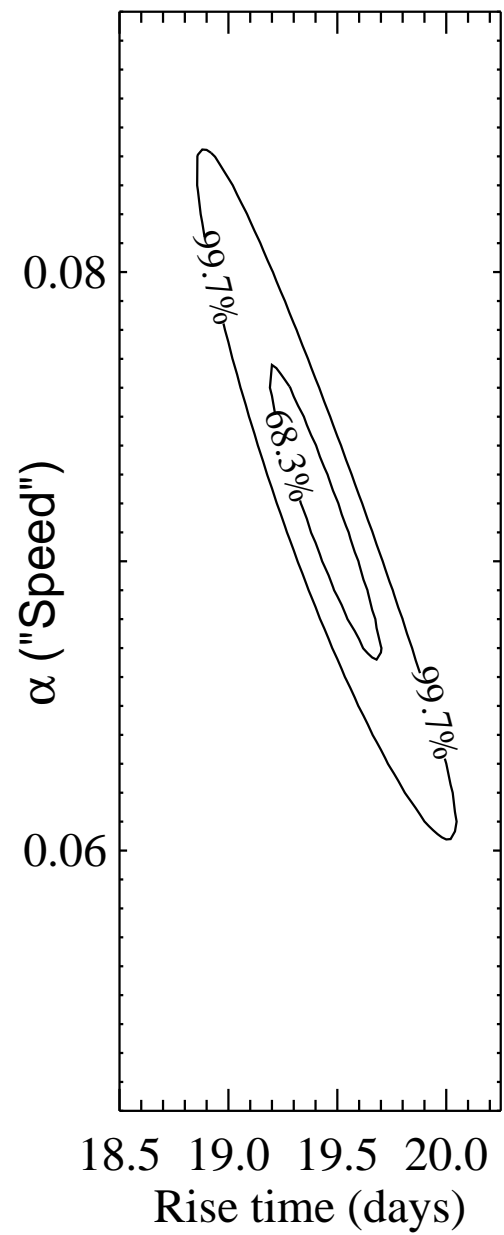
\end{document}